\documentclass[12pt,a4paper,twoside]{article}
\usepackage{latexsym,amsmath,amssymb,amsfonts,amsthm}

\def\nabla{\bigtriangledown}

\newcommand{ \C} {\mbox{\rm I$\!$C}}

\makeindex \textheight 22.5cm
\begin{document}

\title{Noncommutative Symmetries and Stability of\\
Black Ellipsoids in Metric--Affine and String Gravity }
\author{Sergiu I. Vacaru \thanks{%
E-mail address:\ vacaru@fisica.ist.utl.pt, ~~ sergiu$_{-}$vacaru@yahoo.com}%
,\ and Evghenii Gaburov \thanks{
e--mail:\ eg35@leicester.ac.uk \quad}  \and {\small {---}} \\
{\small \textit{* Centro Multidisciplinar de Astrofisica - CENTRA,
Departamento de Fisica,}}\\
{\small \textit{Instituto Superior Tecnico, Av. Rovisco Pais 1, Lisboa,
1049-001, Portugal}} \\
{\small {and}} \\
{\small $\dag$ \ Department of Physics and Astronomy, University}\\
{\small of Leicester, University Road, Leicester, LE1 7RH, UK} \\
}
\date{October 14, 2004}
\maketitle

\begin{abstract}
We construct new classes of exact solutions in metric--affine gravity (MAG)
with string corrections by the antisymmetric $H$--field. The solutions are
parametrized by generic off--diagonal metrics possessing noncommutative
symmetry associated to anholonomy framerelations and related nonlinear
connection (N--connection) structure. We analyze the horizon and geodesic
properties of a class of off--diagonal metrics with deformed spherical
symmetries. The maximal analytic extension of ellipsoid type metrics are
constructed and the Penrose diagrams are analyzed with respect to adapted
frames. We prove that for small deformations (small eccentricities) there
are such metrics that the geodesic behaviour is similar to the Schwarzcshild
one. We conclude that some static and stationary ellipsoid configurations
may describe black ellipsoid objects. The new class of spacetimes do not
possess Killing symmetries even in the limits to the general relativity and,
in consequence, they are not prohibited by black hole uniqueness theorems.
Such static ellipsoid (rotoid) configurations are compatible with the cosmic
cenzorship criteria. We study the perturbations of two classes of static
black ellipsoid solutions of four dimensional gravitational field equations.
The analysis is performed in the approximation of small eccentricity
deformations of the Schwarzschild solution. We conclude that such
anisotropic black hole objects may be stable with respect to the
perturbations parametrized by the Schrodinger equations in the framework of
the one--dimensional inverse scattering theory. We emphasize that the
anholonomic frame method of generating exact solutions is a general one for
off--diagonal metrics (and linear and nonlinear connections) depending on
2-3 variables, in various types of gravity theories.

\vskip5pt.

Pacs:\ 04.50.+h, 04.20.JB, 02.40.-k,

MSC numbers: 83D05, 83C15, 83E15, 53B40, 53C07, 53C60
\end{abstract}

\tableofcontents

\newpage


\section{Introduction}

In the past much effort has been made to construct and investigate exact
solutions of  gravitational field equations with spherical/cylindrical
symmetries and/or with time dependence, paramertrized by metrics
diagonalizable by certain coordinate transforms. Recently, the off--diagonal
metrics were considered in a new manner by diagonalizing them with respect
to anholonomic frames with associated nonlinear connection structure \cite%
{v,v1,vsbd,vspd1}. There were constructed new classes of exact solutions of
Einstein's equations in three (3D), four (4D) and five (5D) dimensions. Such
solutions posses a generic geometric local anisotropy (\textit{e.g.} static
black hole and/or cosmological solutions with ellipsoidal or toroidal
symmetry, various soliton--dilaton 2D and 3D configurations in 4D gravity,
and wormholes and flux tubes with anisotropic polarizations and/or running
constants with different extensions to backgrounds of rotation ellipsoids,
elliptic cylinders, bipolar and toroidal symmetry and anisotropy).

A number of ansatz with off--diagonal metrics were investigated in higher
dimensional gravity (see, for instance, the Salam, Strathee, Percacci and
Randjbar--Daemi works \cite{sal}) which showed that off--diagonal components
in higher dimensional metrics are equi\-valent to including $U(1),SU(2)$ and
$SU(3)$ gauge fields. There are various generalizations of the Kaluza--Klein
gravity when the compactifications of off--diagonal metrics are considered
with the aim to reduce the vacuum 5D gravity to effective Einstein gravity
and Abelian or non--Abelian gauge theories. There were also constructed 4D
exact solutions of Einstein equations with matter fields and cosmological
constants like black torus and black strings induced from some 3D black hole
configurations by considering 4D off--diagonal metrics whose curvature
scalar splits equivalently into a curvature term for a diagonal metric
together with a cosmological constant term and/or a Lagrangian for gauge
(electromagnetic) field \cite{lemos}.

We can model certain effective (diagonal metric) gravitational and matter
fields interactions for some particular off--diagonal metric ansatz and
redefinitions of Lagrangians. However, in general, the vacuum gravitational
dynamics can not be associated to any matter field contributions. This holds
true even if we consider non--Riemanian generalizations from string and/or
metric--affine gravity (MAG) \cite{mag}. In this work (being the third
partner of the papers \cite{vmag1,vmag2}), we prove that such solutions are
not with usual Killing symmetries but admit certain anholonomic
noncommutative symmetries and preserve such properties if the constructions
are extended to MAG and string gravity (see also \cite{vncs} for extensions
to complex and/or noncommutative gravity).

There are constructed the maximal analytic extension of a class of static
metrics with deformed spherical symmetry (containing as particular cases
ellipsoid configurations). We analyze the Penrose diagrams and compare the
results with those for the Reissner--Nordstrom solution. Then we state the
conditions when the geodesic congruence with 'ellipsoid' type symmetry can
be reduced to the Schwarzschild configuration. We argue that in this case we
may generate some static black ellipsoid solutions which, for corresponding
parametrizations of off--diagonal metric coefficients, far away from the
horizon, satisfy the asymptotic conditions of the Minkowski spacetime.

For the new classes of ''off--diagonal'' spacetimes possessing
noncommutative symmetries, we extend the methods elaborated to investigate
the perturbations and stability of black hole metrics. The theory of
perturbations of the Schwarzschild spacetime black holes was initiated in
Ref. \cite{rw}, developed in a series of works, e. g. Refs \cite{vis,fried},
and related \cite{dei} to the theory of inverse scattering and its
ramifications (see, for instanse, Refs. \cite{fad}). The results on the
theory of perturbations and stability of the Schwarzschild,
Reissner--Nordstrom and Kerr solutions are summarized in a monograph \cite%
{chan}. As alternative treatments of the stability of black holes we cite
Ref. \cite{mon}.

Our first aim is to investigate such off--diagonal gravitational
configurations in MAG and string gravity (defined by anholonomic frames with
associated nonlinear connection structure) which describe black hole
solutions with deformed horizons, for instance, with a static ellipsoid
hypersurface. The second aim is to study perturbations of black ellipsoids
and to prove that there are such static ellipsoid like configurations which
are stable with respect to perturbations of a fixed type of anisotropy (i.
e. for certain imposed anholonomic constraints). The main idea of a such
proof is to consider small (ellipsoidal, or another type) deformations of
the Schwarzschild metric and than to apply the already developed methods of
the theory of perturbations of classical black hole solutions, with a
re--definition of the formalism for adapted anholonomic frames.

We note that the solutions defining black ellipsoids are very different from
those defining ellipsoidal shapes in general relativity (see Refs. \cite{es}%
) associated to some perfect--fluid bodies, rotating configurations or to
some families of confocal ellipsoids in Reimannian spaces. Our black
ellipsoid metrics are parametrized by generic off--diagonal ansatz with
anholonomically deformed Killing symmetry and not subjected to uniqueness
theorems. Such ansatz are more general than the class of vacuum solutions
which can not be written in diagonal form \cite{cans} (see details in Refs. %
\cite{vth,vncs}).

The paper is organized as follows: In Sec. 2 we outline the necessary
results on off--diagonal metrics and anhlonomic frames with associated
nonlinear connection structure. We write the system of Einstein--Proca
equations from MAG with string corrections of the antisymmetric $H$--tensor
from bosonic string theory. We introduce a general off--diagonal metric
ansatz and derive the corresponding system of Einstein equations with
anholonomic variables. In Sec. 3 we argue that noncommutative anholonomic
geometries can be associated to real off--diagonal metrics and show two
simple realisations within the algebra for complex matrices. Section 4 is
devoted to the geometry and physics of four dimensional static black
ellipsoids. We illustrate how such solutions can be constructed by using
anholonomic deformations of the Scwarzshild metric, define analytic
extensions of black ellipsoid metrics and analyze the geodesic behaviour of
the static ellipsoid backgrounds. We conclude that black ellipsoid metrics
posses specific noncommutative symmetries. We outline a perturbation theory
of anisotropic black holes and prove the stability of black ellipsoid
objects in Sec. 5. Then, in Sec. 6 we discuss how the method of anholonomic
frame transforms can be related solutions for ellipsoidal shapes and generic
off--diagonal solutions constructed by F. Canfora and H. -J.\ Schmidt. \ We
outline the work and present conclusions in Sec. 7.

There are used the basic notations and conventions stated in Refs. \cite%
{vmag1,vmag2}.

\section{Anholonomic Frames and Off--Diagonal Metrics}

We consider a 4D manifold $V^{3+1}$ (for MAG and string gravity with
possible torsion and nonmetricity structures \cite{mag,vmag1,vmag2}) enabled
with local coordinates $u^{\alpha }=\left( x^{i},y^{a}\right) $ where the
indices of type $i,j,k,...$ run values $1$ and $2$ and the indices $%
a,b,c,... $ take values $3$ and $4;$ $\ y^{3}=v=\varphi $ and $y^{4}=t$ are
considered respectively as the ''anisotropic'' and time like coordinates
(subjected to some constraints). It is supposed that such spacetimes can
also admit nontrivial torsion structures induced by certain frame transforms.

The quadratic line element
\begin{equation}
ds^{2}={g}_{\alpha \beta }\left( x^{i},v\right) du^{\alpha }du^{\beta },
\label{cmetric4}
\end{equation}%
is parametrized by a metric ansatz {%
\begin{equation}
{g}_{\alpha \beta }=\left[
\begin{array}{cccc}
g_{1}+w_{1}^{\ 2}h_{3}+n_{1}^{\ 2}h_{4} & w_{1}w_{2}h_{3}+n_{1}n_{2}h_{4} &
w_{1}h_{3} & n_{1}h_{4} \\
w_{1}w_{2}h_{3}+n_{1}n_{2}h_{4} & g_{2}+w_{2}^{\ 2}h_{3}+n_{2}^{\ 2}h_{4} &
w_{2}h_{3} & n_{2}h_{4} \\
w_{1}h_{3} & w_{2}h_{3} & h_{3} & 0 \\
n_{1}h_{4} & n_{2}h_{4} & 0 & h_{4}%
\end{array}%
\right] ,  \label{ansatzc4}
\end{equation}%
} with $g_{i}=g_{i}\left( x^{i}\right) ,h_{a}=h_{ai}\left( x^{k},v\right) $
and $n_{i}=n_{i}\left( x^{k},v\right) $ being some functions of necessary
smoothly class or even singular in some points and finite regions. The
coefficients $g_{i}$ depend only on ''holonomic'' variables $x^{i}$ but the
rest of coefficients may also depend on one ''anisotropic'' (anholonomic)
variable $y^{3}=v;$ the ansatz does not depend on the time variable $%
y^{4}=t; $ we shall search for static solutions.

The spacetime may be provided with a general anholonomic frame structure of
tetrads, or vierbiends,
\begin{equation}
e_{\alpha }=A_{\alpha }^{\beta }\left( u^{\gamma }\right) \partial /\partial
u^{\beta },  \label{transftet}
\end{equation}%
satisfying some anholonomy \ relations
\begin{equation}
e_{\alpha }e_{\beta }-e_{\beta }e_{\alpha }=w_{\alpha \beta }^{\gamma
}\left( u^{\varepsilon }\right) e_{\gamma },  \label{anhol}
\end{equation}%
where $w_{\alpha \beta }^{\gamma }\left( u^{\varepsilon }\right) $ are
called the coefficients of anholonomy. A 'holonomic' frame, for instance, a
coordinate frame, $e_{\alpha }=\partial _{\alpha }=\partial /\partial
u^{\alpha },$ is defined as a set of tetrads satisfying the holonomy
conditions
\begin{equation*}
\partial _{\alpha }\partial _{\beta }-\partial _{\beta }\partial _{\alpha
}=0.
\end{equation*}

We can 'efectively' diagonalize the line element (\ref{cmetric4}),
\begin{equation}
\delta s^{2}=g_{1}(dx^{1})^{2}+g_{2}(dx^{2})^{2}+h_{3}(\delta
v)^{2}+h_{4}(\delta y^{4})^{2},  \label{dmetric4}
\end{equation}%
with respect to the anholonomic co--frame
\begin{equation}
\delta ^{\alpha }=(d^{i}=dx^{i},\delta ^{a}=dy^{a}+N_{i}^{a}dx^{i})=\left(
d^{i},\delta v=dv+w_{i}dx^{i},\delta y^{4}=dy^{4}+n_{i}dx^{i}\right)
\label{ddif4}
\end{equation}%
which is dual to the anholonomic frame
\begin{equation}
\delta _{\alpha }=(\delta _{i}=\partial _{i}-N_{i}^{a}\partial _{a},\partial
_{b})=\left( \delta _{i}=\partial _{i}-w_{i}\partial _{3}-n_{i}\partial
_{4},\partial _{3},\partial _{4}\right) ,  \label{dder4}
\end{equation}%
where $\partial _{i}=\partial /\partial x^{i}$ and $\partial _{b}=\partial
/\partial y^{b}$ are usual partial derivatives. The tetrads $\delta _{\alpha
}$ and $\delta ^{\alpha }$ are anholonomic because, in general, they satisfy
the anholonomy relations (\ref{anhol}) with some non--trivial coefficients,
\begin{equation}
w_{ij}^{a}=\delta _{i}N_{j}^{a}-\delta
_{j}N_{i}^{a},~w_{ia}^{b}=-~w_{ai}^{b}=\partial _{a}N_{i}^{b}.  \label{anh}
\end{equation}%
The anholonomy is induced by the coefficients $N_{i}^{3}=w_{i}$ and $%
N_{i}^{4}=n_{i}$ which ''elongate'' partial derivatives and differentials if
we are working with respect to anholonomic frames. This results in a more
sophisticate differential and integral calculus (as in the tetradic and/or
spinor gravity), but simplifies substantially the tensor computations,
because we are dealing with diagonalized metrics. In order to construct
exact 'off--diagonal' solutions with 4D metrics depending on 3 variables $%
\left( x^{k},v\right) $ it is more convenient to work with respect to
anholonomic frames (\ref{dder4}) and (\ref{ddif4}) for diagonalized metrics (%
\ref{dmetric4}) than to consider directly the \ ansatz (\ref{cmetric4}) \cite%
{v,v1,vsbd,vspd1}.

There is a 'preferred' linear connection constructed only from the
components $\left( g_{i},h_{a},N_{k}^{b}\right) $, called the canonical
distinguished linear connection, which is similar to the metric connection
introduced by the Christoffel symbols in the case of holonomic bases, i. e.
being constructed only from the metric components and satisfying the
metricity conditions. It is parametrized by the coefficients,\ $\Gamma _{\
\beta \gamma }^{\alpha }=\left( L_{\ jk}^{i},L_{\ bk}^{a},C_{\ jc}^{i},C_{\
bc}^{a}\right) $ stated with respect to the anholonomic frames (\ref{dder4})
and (\ref{ddif4}) as
\begin{eqnarray}
L_{\ jk}^{i} &=&\frac{1}{2}g^{in}\left( \delta _{k}g_{nj}+\delta
_{j}g_{nk}-\delta _{n}g_{jk}\right) ,  \label{dcon} \\
L_{\ bk}^{a} &=&\partial _{b}N_{k}^{a}+\frac{1}{2}h^{ac}\left( \delta
_{k}h_{bc}-h_{dc}\partial _{b}N_{k}^{d}-h_{db}\partial _{c}N_{k}^{d}\right) ,
\notag \\
C_{\ jc}^{i} &=&\frac{1}{2}g^{ik}\partial _{c}g_{jk},\ C_{\ bc}^{a}=\frac{1}{%
2}h^{ad}\left( \partial _{c}h_{db}+\partial _{b}h_{dc}-\partial
_{d}h_{bc}\right) ,  \notag
\end{eqnarray}%
computed for the ansatz (\ref{ansatzc4}). This induces a linear covariant
derivative locally adapted to the nonlinear connection structure
(N--connection, see details, for instance, in Refs. \cite{ma,v,vth}). By
straightforward calculations, we can verify that for $D_{\alpha }$ defined
by $\Gamma _{\ \beta \gamma }^{\alpha }$ with the components (\ref{dcon})
the condition $D_{\alpha }g_{\beta \gamma }=0$ is satisfied.

We note that on (pseudo) Riemannian spaces the N--connection is an object
completely defined by anholonomic frames, when the coefficients of tetradic
transform (\ref{transftet}), $A_{\alpha }^{\beta }\left( u^{\gamma }\right) ,
$ are parametrized explicitly via certain values $\left( N_{i}^{a},\delta
_{i}^{j},\delta _{b}^{a}\right) ,$ where $\delta _{i}^{j}$ $\ $and $\delta
_{b}^{a}$ are the Kronecker symbols. By straightforward calculations we can
compute (see, for instance Ref. \cite{mtw}) that the coefficients of the
Levi--Civita metric connection
\begin{equation*}
\Gamma _{\alpha \beta \gamma }^{\bigtriangledown }=g\left( \delta _{\alpha
},\bigtriangledown _{\gamma }\delta _{\beta }\right) =g_{\alpha \tau }\Gamma
_{\beta \gamma }^{\bigtriangledown \tau },\,
\end{equation*}%
associated to a covariant derivative operator $\bigtriangledown ,$
satisfying the metricity condition $\bigtriangledown _{\gamma }g_{\alpha
\beta }=0$ for $g_{\alpha \beta }=\left( g_{ij},h_{ab}\right) ,$
\begin{equation}
\Gamma _{\alpha \beta \gamma }^{\bigtriangledown }=\frac{1}{2}\left[ \delta
_{\beta }g_{\alpha \gamma }+\delta _{\gamma }g_{\beta \alpha }-\delta
_{\alpha }g_{\gamma \beta }+g_{\alpha \tau }w_{\gamma \beta }^{\tau
}+g_{\beta \tau }w_{\alpha \gamma }^{\tau }-g_{\gamma \tau }w_{\beta \alpha
}^{\tau }\right] ,  \label{lcsym}
\end{equation}%
are given with respect to the anholonomic basis (\ref{ddif4}) by the
coefficients
\begin{equation}
\Gamma _{\beta \gamma }^{\bigtriangledown \tau }=\left( L_{\ jk}^{i},L_{\
bk}^{a}-\frac{\partial N_{k}^{a}}{\partial y^{b}},C_{\ jc}^{i}+\frac{1}{2}%
g^{ik}\Omega _{jk}^{a}h_{ca},C_{\ bc}^{a}\right) ,  \label{lccon}
\end{equation}%
where $\Omega _{jk}^{a}=\delta _{k}N_{j}^{a}-\delta _{j}N_{k}^{a}.$ The
anholonomic frame structure may induce on (pseudo) Riemannian spacetimes
nontrivial torsion structures. For instance, the canonical connection (\ref%
{dcon}), in general, has nonvanishing torsion components
\begin{equation}
T_{ja}^{i}=-T_{aj}^{i}=C_{ja}^{i},T_{jk}^{a}=-T_{kj}^{a}=\Omega
_{kj}^{a},T_{bk}^{a}=-T_{kb}^{a}=\partial _{b}N_{k}^{a}-L_{bk}^{a}.
\label{torsion}
\end{equation}%
This is a ''pure'' anholonomic frame effect. We can conclude that the
Einstein theory transforms into an effective Einstein--Cartan theory with
anholonomically induced torsion if the general relativity is formulated with
respect to general anholonomic frame bases. In this paper we shall also
consider distorsions of the Levi--Civia connection induced by nonmetricity.

A very specific property of off--diagonal metrics is that they can define
different classes of linear connections which satisfy the metricity
conditions for a given metric, or inversely, there is a certain class of
metrics which satisfy the metricity conditions for a given linear
connection. \ This result was originally obtained by A. Kawaguchi \cite{kaw}
(Details can be found in Ref. \cite{ma}, see Theorems 5.4 and 5.5 in Chapter
III, formulated for vector bundles; here we note that similar proofs hold
also on manifolds enabled with anholonomic frames associated to a
N--connection structure.)

The Levi--Civita connection does not play an exclusive role on
non--Riemannian spaces. For instance, the torsion on spaces provided with
N--connection is induced by anholonomy relation and both linear connections (%
\ref{dcon}) and (\ref{lccon}) are compatible with the same metric and
transform into the usual Levi--Civita coefficients for vanishing
N--connection and ''pure'' holonomic coordinates (see related details in
Refs. \cite{vmag1,vmag2}). This means that to an off--diagonal metric we can
associated different covariant differential calculi, all being compatible
with the same metric structure (like in noncommutative geometry, which is
not a surprising \ fact because the anolonomic frames satisfy by definition
some noncommutative relations (\ref{anhol})). In such cases we have to
select a particular type of connection following some physical or
geometrical arguments, or to impose some conditions when there is a single
compatible linear connection constructed only from the metric and
N--coefficients.

The dynamics of generalized Finsler--affine string gravity is defined by the
system of field equations (see Proposition 3.1 in Ref. \cite{vmag2})%
\begin{eqnarray}
\widehat{\mathbf{R}}_{\alpha \beta }-\frac{1}{2}\mathbf{g}_{\alpha \beta }%
\overleftarrow{\mathbf{\hat{R}}} &=&\tilde{\kappa}\left( \mathbf{\Sigma }%
_{\alpha \beta }^{[\phi ]}+\mathbf{\Sigma }_{\alpha \beta }^{[\mathbf{m}]}+%
\mathbf{\Sigma }_{\alpha \beta }^{[\mathbf{T}]}\right) ,  \label{fagfe} \\
\widehat{\mathbf{D}}_{\nu }\mathbf{H}^{\nu \mu } &=&\mu ^{2}\mathbf{\phi }%
^{\mu },  \notag \\
\widehat{\mathbf{D}}^{\nu }\mathbf{H}_{\nu \lambda \rho } &=&0  \notag
\end{eqnarray}%
for
\begin{equation*}
\mathbf{H}_{\nu \lambda \rho }=\widehat{\mathbf{Z}}_{\ \nu \lambda \rho }+%
\widehat{\mathbf{H}}_{\nu \lambda \rho }
\end{equation*}%
being the antisymmetric torsion field
\begin{equation*}
\mathbf{H}_{\nu \lambda \rho }=\delta _{\nu }\mathbf{B}_{\lambda \rho
}+\delta _{\rho }\mathbf{B}_{\nu \lambda }+\delta _{\lambda }\mathbf{B}_{\nu
\rho }
\end{equation*}%
of the antisymmetric $\mathbf{B}_{\lambda \rho }$ in bosonic string theory
(for simplicity, we restrict our considerations to the sigma model with $H$%
--field corrections and zero dilatonic field). The covariant derivative $%
\widehat{\mathbf{D}}_{\nu }$ is defined by the coefficients (\ref{dcon}) (we
use in our references the ''boldfaced'' indices when it is necessary to
emphasize that the spacetime is provided with N--connection structure. The
distorsion $\widehat{\mathbf{Z}}_{\ \nu \lambda \rho }$ of the Levi--Civita
connection, when
\begin{equation*}
\Gamma _{\beta \gamma }^{\tau }=\Gamma _{\bigtriangledown \beta \gamma
}^{\tau }+\widehat{\mathbf{Z}}_{\beta \gamma }^{\tau },
\end{equation*}%
from (\ref{fagfe}) is defined by the torsion $\widehat{\mathbf{T}}$ with the
components computed for $\widehat{\mathbf{D}}$ by applying the formulas (\ref%
{torsion}),
\begin{equation*}
\widehat{\mathbf{Z}}_{\alpha \beta }=\delta _{\beta }\rfloor \widehat{%
\mathbf{T}}_{\alpha }-\delta _{\alpha }\rfloor \widehat{\mathbf{T}}_{\beta }+%
\frac{1}{2}\left( \delta _{\alpha }\rfloor \delta _{\beta }\rfloor \widehat{%
\mathbf{T}}_{\gamma }\right) \delta ^{\gamma },
\end{equation*}%
see Refs. \cite{vmag1,vmag2} on definition of the interior product ''$%
\rfloor $'' and differential forms like $\widehat{\mathbf{T}}_{\beta }$ on
spaces provided with N--connection structure. The tensor $\mathbf{H}_{\nu
\mu }\doteqdot \widehat{\mathbf{D}}_{\nu }\mathbf{\phi }_{\mu }-\widehat{%
\mathbf{D}}_{\mu }\mathbf{\phi }_{\nu }+w_{\mu \nu }^{\gamma }\mathbf{\phi }%
_{\gamma }$ is the field strengths of the Abelian Proca field $\mathbf{\phi }%
^{\alpha },$ with $\mu ,\tilde{\kappa}=const,$
\begin{equation*}
\mathbf{\Sigma }_{\alpha \beta }^{[\phi ]}=\mathbf{H}_{\alpha }^{\ \mu }%
\mathbf{H}_{\beta \mu }-\frac{1}{4}\mathbf{g}_{\alpha \beta }\mathbf{H}_{\mu
\nu }^{\ }\mathbf{H}^{\mu \nu }+\mu ^{2}\mathbf{\phi }_{\alpha }\mathbf{\phi
}_{\beta }-\frac{\mu ^{2}}{2}\mathbf{g}_{\alpha \beta }\mathbf{\phi }_{\mu }%
\mathbf{\phi }^{\mu },
\end{equation*}%
where the source%
\begin{equation*}
\mathbf{\Sigma }_{\alpha \beta }^{[\mathbf{T}]}=\mathbf{\Sigma }_{\alpha
\beta }^{[\mathbf{T}]}\left( \widehat{\mathbf{T}},\mathbf{H}_{\nu \lambda
\rho }\right)
\end{equation*}%
contains contributions of the torsion fields $\widehat{\mathbf{T}}$ and $%
\mathbf{H}_{\nu \lambda \rho }.$Thefield $\mathbf{\phi }_{\alpha }$ certain
irreducible components of torsion and nonmetricity in MAG, see  \cite{mag}
and Theorem 3.2 in  \cite{vmag2}.

Our aim is to elaborate a method of constructing exact solutions of
equations (\ref{fagfe}) for vanishing matter fields, $\mathbf{\Sigma }%
_{\alpha \beta }^{[\mathbf{m}]}=0.$ The ansatz for the field $\mathbf{\phi }%
_{\mu }$ is taken in the form
\begin{equation*}
\mathbf{\phi }_{\mu }=\left[ \mathbf{\phi }_{i}\left( x^{k}\right) ,\mathbf{%
\phi }_{a}=0\right]
\end{equation*}%
for $i,j,k...=1,2$ and $a,b,...=3,4.$ The Proca equations $\widehat{\mathbf{D%
}}_{\nu }\mathbf{H}^{\nu \mu }=\mu ^{2}\mathbf{\phi }^{\mu }$ for $\mu
\rightarrow 0$ (for simplicity) transform into
\begin{equation}
\partial _{1}\left[ \left( g_{1}\right) ^{-1}\partial ^{1}\mathbf{\phi }_{k}%
\right] +\partial _{2}\left[ \left( g_{2}\right) ^{-1}\partial ^{2}\mathbf{%
\phi }_{k}\right] =L_{ki}^{j}\partial ^{i}\mathbf{\phi }_{j}-L_{ij}^{i}%
\partial ^{j}\mathbf{\phi }_{k}.  \label{ans21}
\end{equation}%
Two examples of solutions of this equation are considered in Ref. \cite%
{vmag2}. In this paper, we do not state any particular configurations and
consider that it is possible always to defie certain $\mathbf{\phi }%
_{i}\left( x^{k}\right) $ satisfying the wave equation (\ref{ans21}). The
energy--momentum tensor $\mathbf{\Sigma }_{\alpha \beta }^{[\phi ]}$ is
computed for one nontrivial value
\begin{equation*}
H_{12}=\partial _{1}\mathbf{\phi }_{2}-\partial _{2}\mathbf{\phi }_{1}.
\end{equation*}
In result, we can represent the source of the fields $\mathbf{\phi }_{k}$ as
\begin{equation*}
\mathbf{\Sigma }_{\alpha \beta }^{[\phi ]}=\left[ \Psi _{2}\left(
H_{12},x^{k}\right) ,\Psi _{2}\left( H_{12},x^{k}\right) ,0,0\right] .
\end{equation*}

The ansatz for the $H$--field is taken in the form
\begin{equation}
\mathbf{H}_{\nu \lambda \rho }=\widehat{\mathbf{Z}}_{\ \nu \lambda \rho }+%
\widehat{\mathbf{H}}_{\nu \lambda \rho }=\lambda _{\lbrack H]}\sqrt{|\mathbf{%
g}_{\alpha \beta }|}\varepsilon _{\nu \lambda \rho }  \label{cond03}
\end{equation}%
where $\varepsilon _{\nu \lambda \rho }$ is completely antisymmetric and $%
\lambda _{\lbrack H]}=const.$ This ansatz satisfies the field equations $%
\widehat{\mathbf{D}}^{\nu }\mathbf{H}_{\nu \lambda \rho }=0$ because the
metric $\mathbf{g}_{\alpha \beta }$ is compatible with $\widehat{\mathbf{D}}.
$ The values $\widehat{\mathbf{H}}_{\nu \lambda \rho }$ have to be defined
in a form to satisfy the condition (\ref{cond03})\ for any $\widehat{\mathbf{%
Z}}_{\ \nu \lambda \rho }$ derived from $\mathbf{g}_{\alpha \beta }$ and, as
a consequence, from (\ref{dcon}) and (\ref{torsion}), for instanese, to
compute them as
\begin{equation*}
\widehat{\mathbf{H}}_{\nu \lambda \rho }=\lambda _{\lbrack H]}\sqrt{|\mathbf{%
g}_{\alpha \beta }|}\varepsilon _{\nu \lambda \rho }-\widehat{\mathbf{Z}}_{\
\nu \lambda \rho }
\end{equation*}%
for defeined values of $\widehat{\mathbf{Z}}_{\ \nu \lambda \rho },\lambda
_{\lbrack H]}$ and $\mathbf{g}_{\alpha \beta }.$ In result, we obtain the
effective energy--momentum tensor in the form%
\begin{equation}
\mathbf{\Sigma }_{\alpha }^{[\phi ]\beta }+\mathbf{\Sigma }_{\alpha
}^{[H]\beta }=\left[ \Upsilon _{2}\left( x^{k}\right) +\frac{\lambda
_{\lbrack H]}^{2}}{4},\Upsilon _{2}\left( x^{k}\right) +\frac{\lambda
_{\lbrack H]}^{2}}{4},\frac{\lambda _{\lbrack H]}^{2}}{4},\frac{\lambda
_{\lbrack H]}^{2}}{4}\right] .  \label{sours01}
\end{equation}

For the source (\ref{sours01}), the system of field equations (\ref{fagfe})
defined for the metric (\ref{dmetric4}) and connection (\ref{dcon}), with
respect to anholonomic frames (\ref{ddif4}) and (\ref{dder4}), transform
into a system of partial differential equations with anholonomic variables %
\cite{v,v1,vth}, see also details in the section 5.3 in Ref. \cite{vmag2},
\begin{eqnarray}
R_{1}^{1}=R_{2}^{2}=-\frac{1}{2g_{1}g_{2}}[g_{2}^{\bullet \bullet }-\frac{%
g_{1}^{\bullet }g_{2}^{\bullet }}{2g_{1}}-\frac{(g_{2}^{\bullet })^{2}}{%
2g_{2}}+g_{1}^{^{\prime \prime }}-\frac{g_{1}^{^{\prime }}g_{2}^{^{\prime }}%
}{2g_{2}}-\frac{(g_{1}^{^{\prime }})^{2}}{2g_{1}}] &=&-\frac{\lambda
_{\lbrack H]}^{2}}{4},  \label{ricci1a} \\
R_{3}^{3}=R_{4}^{4}=-\frac{\beta }{2h_{3}h_{4}}=-\frac{1}{2h_{3}h_{4}}\left[
h_{4}^{\ast \ast }-h_{4}^{\ast }\left( \ln \sqrt{|h_{3}h_{4}|}\right) ^{\ast
}\right]  &=&-\frac{\lambda _{\lbrack H]}^{2}}{4}-\Upsilon _{2}\left(
x^{k}\right) ,  \label{ricci2a} \\
R_{3i}=-w_{i}\frac{\beta }{2h_{4}}-\frac{\alpha _{i}}{2h_{4}} &=&0,
\label{ricci3a} \\
R_{4i}=-\frac{h_{4}}{2h_{3}}\left[ n_{i}^{\ast \ast }+\gamma n_{i}^{\ast }%
\right]  &=&0,  \label{ricci4a}
\end{eqnarray}%
where
\begin{equation}
\alpha _{i}=\partial _{i}h_{4}^{\ast }-h_{4}^{\ast }\partial _{i}\ln \sqrt{%
|h_{3}h_{4}|},~\gamma =3h_{4}^{\ast }/2h_{4}-h_{3}^{\ast }/h_{3},
\label{abc}
\end{equation}%
and the partial derivatives are denoted $g_{1}^{\bullet }=\partial
g_{1}/\partial x^{1},g_{1}^{^{\prime }}=\partial g_{1}/\partial x^{2}$ and $%
h_{3}^{\ast }=\partial h_{3}/\partial v.$ We \ can additionally impose the
condition $\delta _{i}N_{j}^{a}=\delta _{j}N_{i}^{a}$ in order to have $%
\Omega _{jk}^{a}=0$ which may be satisfied, for instance, if
\begin{equation*}
w_{1}=w_{1}\left( x^{1},v\right) ,n_{1}=n_{1}\left( x^{1},v\right)
,w_{2}=n_{2}=0,
\end{equation*}%
or, inversely, if
\begin{equation*}
w_{1}=n_{1}=0,w_{2}=w_{2}\left( x^{2},v\right) ,n_{2}=n_{2}\left(
x^{2},v\right) .
\end{equation*}%
In this paper we shall select a class of static solutions parametrized by
the conditions
\begin{equation}
w_{1}=w_{2}=n_{2}=0.  \label{cond1}
\end{equation}

\ The system of equations (\ref{ricci1a})--(\ref{ricci4a}) can be integrated
in general form \cite{vmag2,vth}. Physical solutions are selected following
some additional boundary conditions, imposed types of symmetries,
nonlinearities and singular behaviour and compatibility in the locally
anisotropic limits with some well known exact solutions.

Finally, we note that there is a difference between our approach and the
so--called ''tetradic'' gravity (see basic details and references in \cite%
{mtw}) when the metric coefficients $g_{\alpha \beta }\left( u^{\gamma
}\right) $ are substituted by tetradic fields $e_{\alpha }^{\underline{%
\alpha }}\left( u^{\gamma }\right) ,$ mutually related by formula $g_{\alpha
\beta }=e_{\alpha }^{\underline{\alpha }}e_{\beta }^{\underline{\beta }}\eta
_{\underline{\alpha }\underline{\beta }}$ with $\eta _{\underline{\alpha }%
\underline{\beta }}$ chosen, for instance, to be the Minkowski metric. In
our case we partially preserve some metric dynamics given by diagonal
effective metric coefficients $\left( g_{i},h_{a}\right) $ but also adapt
the calculus to tetrads respectively defied by $\left( N_{i}^{a},\delta
_{i}^{j},\delta _{b}^{a}\right) ,$ see (\ref{ddif4}) and (\ref{dder4}). This
substantially simplifies the method of constructing exact solutions and also
reflects new type symmetries of such classes of metrics.

\section{Anholonomic Noncommutative Symmetries}

\label{ncgg}

The nontrivial anholonomy coefficients, see (\ref{anhol}) and (\ref{anh})
induced by off--diagonal metric (\ref{cmetric4}) (and associated
N--connection) coefficients emphasize a kind of Lie algebra noncommutativity
relation. \ In this section, we analyze a simple realizations of
noncommutative geometry of anholonomic frames within the algebra of complex $%
k\times k$ matrices, $M_{k}(\C,u^{\alpha })$ depending on coordinates $%
u^{\alpha }$ on \ spacetime $V^{n+m}$ connected to complex Lie algebras $%
SL\left( k,\C\right) $ (see Ref. \cite{vncs} for similar constructions with
the group $SU_{k}).$

We consider matrix valued functions of necessary smoothly class derived from
the anholonomic frame relations (\ref{anhol}) (being similar to the Lie
algebra relations) with the coefficients (\ref{anh}) induced by
off--diagonal metric terms in (\ref{ansatzc4}) and by N--connection
coefficients $N_{i}^{a}.$ We use algebras of complex matrices in order to
have the possibility for some extensions to complex solutions and to relate
the constructions to noncommutative/complex gravity). For commutative
gravity models, the anholonomy coefficients $w_{~\alpha \beta }^{\gamma }$
are real functions but there are considered also complex metrics and tetrads
related to noncommutative gravity \cite{ncg}.

Let us consider the basic relations for the simplest model of noncommutative
geometry realized with the algebra of complex $\left( k\times k\right) $
noncommutative matrices \cite{dub}, $M_{k}(\C).$ Any element $M\in M_{k}(\C)$
can be represented as a linear combination of the unit $\left( k\times
k\right) $ matrix $I$ and $\left( k^{2}-1\right) $ hermitian traseless
matrices $q_{\underline{\alpha }}$ with the underlined index $\underline{%
\alpha }$ running values $1,2,...,k^{2}-1,$ i. e.
\begin{equation*}
M=\alpha \ I+\sum \beta ^{\underline{\alpha }}q_{\underline{\alpha }}
\end{equation*}%
for some constants $\alpha $ and $\beta ^{\underline{\alpha }}.$ It is
possible to chose the basis matrices $q_{\underline{\alpha }}$ satisfying
the relations%
\begin{equation}
q_{\underline{\alpha }}q_{\underline{\beta }}=\frac{1}{k}\rho _{\underline{%
\alpha }\underline{\beta }}I+Q_{\underline{\alpha }\underline{\beta }}^{%
\underline{\gamma }}q_{\underline{\gamma }}-\frac{i}{2}f_{~\underline{\alpha
}\underline{\beta }}^{\underline{\gamma }}q_{\underline{\gamma }},
\label{gr1}
\end{equation}%
where $i^{2}=-1$ and the real coefficients satisfy the properties
\begin{equation*}
Q_{\underline{\alpha }\underline{\beta }}^{\underline{\gamma }}=Q_{%
\underline{\beta }\underline{\alpha }}^{\underline{\gamma }},\ Q_{\underline{%
\gamma }\underline{\beta }}^{\underline{\gamma }}=0,\ f_{~\underline{\alpha }%
\underline{\beta }}^{\underline{\gamma }}=-f_{~\underline{\beta }\underline{%
\alpha }}^{\underline{\gamma }},f_{~\underline{\gamma }\underline{\alpha }}^{%
\underline{\gamma }}=0
\end{equation*}%
with $f_{~\underline{\alpha }\underline{\beta }}^{\underline{\gamma }}$
being the structure constants of the Lie group $SL\left( k,\C\right) $ and
the Killing--Cartan metric tensor $\rho _{\underline{\alpha }\underline{%
\beta }}=f_{~\underline{\alpha }\underline{\gamma }}^{\underline{\tau }}f_{~%
\underline{\tau }\underline{\beta }}^{\underline{\gamma }}.$ This algebra
admits a formalism of interior derivatives $\widehat{\partial }_{\underline{%
\gamma }}$ defied by relations
\begin{equation}
\widehat{\partial }_{\underline{\gamma }}q_{\underline{\beta }}=ad\left( iq_{%
\underline{\gamma }}\right) q_{\underline{\beta }}=i[q_{\underline{\gamma }%
},q_{\underline{\beta }}]=f_{~\underline{\gamma }\underline{\beta }}^{%
\underline{\alpha }}q_{\underline{\alpha }}  \label{der1a}
\end{equation}
and%
\begin{equation}
\widehat{\partial }_{\underline{\alpha }}\widehat{\partial }_{\underline{%
\beta }}-\widehat{\partial }_{\underline{\beta }}\widehat{\partial }_{%
\underline{\alpha }}=f_{~\underline{\alpha }\underline{\beta }}^{\underline{%
\gamma }}\widehat{\partial }_{\underline{\gamma }}  \label{jacob}
\end{equation}%
(the last relation follows the Jacoby identity and is quite similar to (\ref%
{anhol}) but with constant values $f_{~\underline{\alpha }\underline{\beta }%
}^{\underline{\gamma }}).$

Our idea is to associate a noncommutative geometry starting from the
anholonomy relations of frames (\ref{anhol}) by adding to the structure
constants $f_{~\underline{\alpha }\underline{\beta }}^{\underline{\gamma }}$
the anhlonomy coefficients $w_{\ \alpha \gamma }^{[N]\tau }$ (\ref{anh}) (we
shall put the label [N] if would be necessary to emphasize that the
anholonomic coefficients are induced by a nolinear connection. Such deformed
structure constants consist from N--connection coefficients $N_{i}^{a}$ and
their first partial derivatives, i. e. they are induced by some
off--diagonal terms in the metric (\ref{ansatzc4}) being a solution of the
gravitational field equations.

We emphasize that there is a rough analogy between formulas (\ref{jacob})
and (\ref{anhol}) because the anholonomy coefficients do not satisfy, in
general, the condition $w_{\ \tau \alpha }^{[N]\tau }=0.$ There is also
another substantial difference because the anholonomy relations are defined
for a manifold of dimension $n+m,$ with Greek indices $\alpha ,\beta ,...$
running values from $1$ to $\ n+m$ but the matrix noncommutativity relations
are stated for traseless matrices labeled by underlined indices $\underline{%
\alpha },\underline{\beta },$ running values from $1$ to $k^{2}-1.$ It is
not possible to satisfy the condition $k^{2}-1=n+m$ by using integer numbers
for arbitrary $n+m.$ We may extend the dimension of spacetime from $n+m$ to
any $n^{\prime }\geq n$ and $m^{\prime }\geq m$ when the condition $%
k^{2}-1=n^{\prime }+m^{\prime }$ can be satisfied by a trivial embedding of
the metric (\ref{ansatzc4}) into higher dimension, for instance, by adding
the necessary number of unities on the diagonal and writing
\begin{equation*}
\widehat{g}_{\underline{\alpha }\underline{\beta }}=\left[
\begin{array}{ccccc}
1 & ... & 0 & 0 & 0 \\
... & ... & ... & ... & ... \\
0 & ... & 1 & 0 & 0 \\
0 & ... & 0 & g_{ij}+N_{i}^{a}N_{j}^{b}h_{ab} & N_{j}^{e}h_{ae} \\
0 & ... & 0 & N_{i}^{e}h_{be} & h_{ab}%
\end{array}%
\right]
\end{equation*}%
and $e_{\underline{\alpha }}^{[N]}=\delta _{\underline{\alpha }}=\left(
1,1,...,e_{\alpha }^{[N]}\right) .$ For simplicity, we preserve the same
type of underlined Greek indices, $\underline{\alpha },\underline{\beta }%
...=1,2,...,k^{2}-1=n^{\prime }+m^{\prime }.$

The anholonomy coefficients $w_{~\alpha \beta }^{[N]\gamma }$ can be
extended with some trivial zero components and for consistency we rewrite
them without labeled indices, $w_{~\alpha \beta }^{[N]\gamma }\rightarrow
W_{~\underline{\alpha }\underline{\beta }}^{\underline{\gamma }}.$ The set
of anholonomy coefficients $w_{~\alpha \beta }^{[N]\gamma }$(\ref{anhol})
may result in degenerated matrices, for instance for certain classes of
exact solutions of the Einstein equations. So, it  would not be a well
defined construction if we shall substitute the structure Lie algebra
constants directly by $w_{~\alpha \beta }^{[N]\gamma }.$ We can consider a
simple extension $w_{~\alpha \beta }^{[N]\gamma }\rightarrow W_{~\underline{%
\alpha }\underline{\beta }}^{\underline{\gamma }}$ when the coefficients $%
w_{~\underline{\alpha }\underline{\beta }}^{\underline{\gamma }}(u^{%
\underline{\tau }})$ for any fixed value $u^{\underline{\tau }}=u_{[0]}^{%
\underline{\tau }}$ would be some deformations of the\ structure constants
of the Lie algebra $SL\left( k,\C\right) ,$ like
\begin{equation}
W_{~\underline{\alpha }\underline{\beta }}^{\underline{\gamma }}=f_{~%
\underline{\alpha }\underline{\beta }}^{\underline{\gamma }}+w_{~\underline{%
\alpha }\underline{\beta }}^{\underline{\gamma }},  \label{anhb}
\end{equation}%
being nondegenerate. \

Instead of the matrix algebra $M_{k}(\C),$ constructed from constant complex
elements, we have also to introduce dependencies on coordinates $u^{%
\underline{\alpha }}=\left( 0,...,u^{\alpha }\right) ,$ for instance, like a
trivial matrix bundle on $V^{n^{\prime }+m^{\prime }},$ and denote this
space $M_{k}(\C,u^{\underline{\alpha }}).$ Any element $B\left( u^{%
\underline{\alpha }}\right) \in M_{k}(\C,u^{\underline{\alpha }})$ with a
noncommutative structure induced by $W_{~\underline{\alpha }\underline{\beta
}}^{\underline{\gamma }}$ is represented as a linear combination of the unit
$(n^{\prime }+m^{\prime })\times (n^{\prime }+m^{\prime })$ matrix $I$ and
the $[(n^{\prime }+m^{\prime })^{2}-1]$ hermitian traceless matrices $q_{%
\underline{\alpha }}\left( u^{\underline{\tau }}\right) $ with the
underlined index $\underline{\alpha }$ running values $1,2,...,(n^{\prime
}+m^{\prime })^{2}-1,$%
\begin{equation*}
B\left( u^{\underline{\tau }}\right) =\alpha \left( u^{\underline{\tau }%
}\right) \ I+\sum \beta ^{\underline{\alpha }}\left( u^{\underline{\tau }%
}\right) q_{\underline{\alpha }}\left( u^{\underline{\tau }}\right)
\end{equation*}%
under condition that the following relation holds:%
\begin{equation*}
q_{\underline{\alpha }}\left( u^{\underline{\tau }}\right) q_{\underline{%
\beta }}\left( u^{\underline{\gamma }}\right) =\frac{1}{n^{\prime
}+m^{\prime }}\rho _{\underline{\alpha }\underline{\beta }}\left( u^{%
\underline{\nu }}\right) +Q_{\underline{\alpha }\underline{\beta }}^{%
\underline{\gamma }}q_{\underline{\gamma }}\left( u^{\underline{\mu }%
}\right) -\frac{i}{2}W_{~\underline{\alpha }\underline{\beta }}^{\underline{%
\gamma }}q_{\underline{\gamma }}\left( u^{\underline{\mu }}\right)
\end{equation*}%
with the same values of $Q_{\underline{\alpha }\underline{\beta }}^{%
\underline{\gamma }}$ from the Lie algebra for $SL\left( k,\C\right) $ but
with the Killing--Cartan like metric tensor defined by anholonomy
coefficients, i. e. $\rho _{\underline{\alpha }\underline{\beta }}\left( u^{%
\underline{\nu }}\right) =W_{~\underline{\alpha }\underline{\gamma }}^{%
\underline{\tau }}\left( u^{\underline{\alpha }}\right) W_{~\underline{\tau }%
\underline{\beta }}^{\underline{\gamma }}\left( u^{\underline{\alpha }%
}\right) .$ For complex spacetimes, we shall consider that the coefficients $%
N_{\underline{i}}^{\underline{a}}$ and $W_{~\underline{\alpha }\underline{%
\beta }}^{\underline{\gamma }}$ may be some complex valued functions of
necessary smooth (in general, with complex variables) class. In result, the
Killing--Cartan like metric tensor $\rho _{\underline{\alpha }\underline{%
\beta }}$ can be also complex.

We rewrite (\ref{anhol}) as
\begin{equation}
e_{\underline{\alpha }}e_{\underline{\beta }}-e_{\underline{\beta }}e_{%
\underline{\alpha }}=W_{~\underline{\alpha }\underline{\beta }}^{\underline{%
\gamma }}e_{\underline{\gamma }}  \label{anh1}
\end{equation}%
being equivalent to (\ref{jacob}) and defining a noncommutative anholonomic
structure (for simplicity, we use the same symbols $e_{\underline{\alpha }}$
as for some 'N--elongated' partial derivatives, but with underlined
indices). The analogs of derivation operators (\ref{der1a}) are stated by
using $W_{~\underline{\alpha }\underline{\beta }}^{\underline{\gamma }},$%
\begin{equation}
e_{\underline{\alpha }}q_{\underline{\beta }}\left( u^{\underline{\gamma }%
}\right) =ad\left[ iq_{\underline{\alpha }}\left( u^{\underline{\gamma }%
}\right) \right] q_{\underline{\beta }}\left( u^{\underline{\gamma }}\right)
=i\left[ q_{\underline{\alpha }}\left( u^{\underline{\gamma }}\right) q_{%
\underline{\beta }}\left( u^{\underline{\gamma }}\right) \right] =W_{~%
\underline{\alpha }\underline{\beta }}^{\underline{\gamma }}q_{\underline{%
\gamma }}  \label{nder1}
\end{equation}

The operators (\ref{nder1}) define a linear space of anholonomic derivations
satisfying the conditions (\ref{anh1}), denoted $AderM_{k}(\C,u^{\underline{%
\alpha }}),$ elongated by N--connection and distinguished into irreducible
h-- and v--components, respectively, into $e_{\underline{i}}$ and $e_{%
\underline{b}},$ like $e_{\underline{\alpha }}=\left( e_{\underline{i}%
}=\partial _{\underline{i}}-N_{\underline{i}}^{\underline{a}}e_{\underline{a}%
},e_{\underline{b}}=\partial _{\underline{b}}\right) .$ The space $AderM_{k}(%
\C,u^{\underline{\alpha }})$ is not a left module over \ the algebra $M_{k}(%
\C,u^{\underline{\alpha }})$ which means that there is a a substantial
difference with respect to the usual commutative differential geometry where
a vector field multiplied on the left by a function produces a new vector
field.

The duals to operators (\ref{nder1}), $e^{\underline{\mu }},$ found from $e^{%
\underline{\mu }}\left( e_{\underline{\alpha }}\right) =\delta _{_{%
\underline{\alpha }}}^{\underline{\mu }}I,$ define a canonical basis of
1--forms $e^{\underline{\mu }}$ connected to the N--connection structure. By
using these forms, we can span a left module over $M_{k}(\C,u^{\underline{%
\alpha }})$ following $q_{\underline{\alpha }}e^{\underline{\mu }}\left( e_{%
\underline{\beta }}\right) =q_{\underline{\alpha }}\delta _{_{\underline{%
\beta }}}^{\underline{\mu }}I=q_{\underline{\alpha }}\delta _{_{\underline{%
\beta }}}^{\underline{\mu }}.$ \ For an arbitrary vector field
\begin{equation*}
Y=Y^{\alpha }e_{\alpha }\rightarrow Y^{\underline{\alpha }}e_{\underline{%
\alpha }}=Y^{\underline{i}}e_{\underline{i}}+Y^{\underline{a}}e_{\underline{a%
}},
\end{equation*}%
it is possible to define an exterior differential (in our case being
N--elongated), starting with the action on a function $\varphi $
(equivalent, a 0--form),
\begin{equation*}
\delta \ \varphi \left( Y\right) =Y\varphi =Y^{\underline{i}}\delta _{%
\underline{i}}\varphi +Y^{\underline{a}}\partial _{\underline{a}}\varphi
\end{equation*}%
when%
\begin{equation*}
\left( \delta \ I\right) \left( e_{\underline{\alpha }}\right) =e_{%
\underline{\alpha }}I=ad\left( ie_{\underline{\alpha }}\right) I=i\left[ e_{%
\underline{\alpha }},I\right] =0,\mbox{ i. e. }\delta I=0,
\end{equation*}%
and
\begin{equation}
\delta q_{\underline{\mu }}(e_{\underline{\alpha }})=e_{\underline{\alpha }%
}(e_{\underline{\mu }})=i[e_{\underline{\mu }},e_{\underline{\alpha }}]=W_{~%
\underline{\alpha }\underline{\mu }}^{\underline{\gamma }}e_{\underline{%
\gamma }}.  \label{aux1}
\end{equation}%
Considering the nondegenerated case, we can invert (\ref{aux1}) as to obtain
a similar expression with respect to $e^{\underline{\mu }},$%
\begin{equation}
\delta (e_{\underline{\alpha }})=W_{~\underline{\alpha }\underline{\mu }}^{%
\underline{\gamma }}e_{\underline{\gamma }}e^{\underline{\mu }},
\label{aux2}
\end{equation}%
from which a very important property follows by using the Jacobi identity, $%
\delta ^{2}=0,$ resulting in a possibility to define a usual Grassman
algebra of $p$--forms with the wedge product $\wedge $ stated as%
\begin{equation*}
e^{\underline{\mu }}\wedge e^{\underline{\nu }}=\frac{1}{2}\left( e^{%
\underline{\mu }}\otimes e^{\underline{\nu }}-e^{\underline{\nu }}\otimes e^{%
\underline{\mu }}\right) .
\end{equation*}%
We can write (\ref{aux2}) as
\begin{equation*}
\delta (e^{\underline{\alpha }})=-\frac{1}{2}W_{~\underline{\beta }%
\underline{\mu }}^{\underline{\alpha }}e^{\underline{\beta }}e^{\underline{%
\mu }}
\end{equation*}%
and introduce the canonical 1--form $e=q_{\underline{\alpha }}e^{\underline{%
\alpha }}$ being coordinate--independent and adapted to the N--connection
structure and satisfying the condition $\delta e+e\wedge e=0.$

In a standard manner, we can introduce the volume element induced by the
canonical Cartan--Killing metric and the corresponding star operator $\star $%
\ (Hodge duality).\ We define the volume element $\sigma $ by using the
complete antisymmetric tensor $\epsilon _{\underline{\alpha }_{1}\underline{%
\alpha }_{2}...\underline{\alpha }_{k^{2}-1}}$as
\begin{equation*}
\sigma =\frac{1}{\left[ (n^{\prime }+m^{\prime })^{2}-1\right] !}\epsilon _{%
\underline{\alpha }_{1}\underline{\alpha }_{2}...\underline{\alpha }%
_{n^{\prime }+m^{\prime }}}e^{\underline{\alpha }_{1}}\wedge e^{\underline{%
\alpha }_{2}}\wedge ...\wedge e^{\underline{\alpha }_{n^{\prime }+m^{\prime
}}}
\end{equation*}%
to which any $\left( k^{2}-1\right) $--form is proportional $\left(
k^{2}-1=n^{\prime }+m^{\prime }\right) .$ The integral of such a form is
defined as the trace of the matrix coefficient in the from $\sigma $ and the
scalar product introduced for any couple of $p$--forms $\varpi $ and $\psi $%
\begin{equation*}
\left( \varpi ,\psi \right) =\int \left( \varpi \wedge \star \psi \right) .
\end{equation*}

Let us analyze how a noncommutative differential form calculus (induced by
an anholonomic structure) can be developed and related to the Hamiltonian
classical and quantum mechanics and Poisson bracket formalism:

For a $p$--form $\varpi ^{\lbrack p]},$ the anti--derivation $i_{Y}$ with
respect to a vector field $Y\in AderM_{k}(\C,u^{\underline{\alpha }})$ can
be defined as in the usual formalism,%
\begin{equation*}
i_{Y}\varpi ^{\lbrack p]}\left( X_{1},X_{2},...,X_{p-1}\right) =\varpi
^{\lbrack p]}\left( Y,X_{1},X_{2},...,X_{p-1}\right)
\end{equation*}%
where $X_{1},X_{2},...,X_{p-1}\in AderM_{k}(\C,u^{\underline{\alpha }}).$ By
a straightforward calculus we can check that for a 2--form $\Xi =\delta $ $e$
one holds
\begin{equation*}
\delta \Xi =\delta ^{2}e=0\mbox{ and }L_{Y}\Xi =0
\end{equation*}%
where the Lie derivative of forms is defined as $L_{Y}\varpi ^{\lbrack
p]}=\left( i_{Y}\ \delta +\delta \ i_{Y}\right) \varpi ^{\lbrack p]}.$

The Hamiltonian vector field $H_{[\varphi ]}$ of an element of algebra $%
\varphi \in M_{k}(\C,u^{\underline{\alpha }})$ is introduced following the
equality $\Xi \left( H_{[\varphi ]},Y\right) =Y\varphi $ which holds for any
vector field. Then, we can define the Poisson bracket of two functions (in a
quantum variant, observables) $\varphi $ and $\chi ,$ $\{\varphi ,\chi
\}=\Xi \left( H_{[\varphi ]},H_{[\chi ]}\right) $ when
\begin{equation*}
\{e_{\underline{\alpha }},e_{\underline{\beta }}\}=\Xi \left( e_{\underline{%
\alpha }},e_{\underline{\beta }}\right) =i[e_{\underline{\alpha }},e_{%
\underline{\beta }}].
\end{equation*}%
This way, a simple version of noncommutative classical and quantum mechanics
(up to a factor like the Plank constant, $\hbar $) is proposed, being
derived by anholonomic relations for a certain class of exact
'off--diagonal' solutions in commutative gravity.

In order to define the algebra of forms $\Omega ^{\ast }\left[ M_{k}(\C,u^{%
\underline{\alpha }})\right] $ over $M_{k}(\C,u^{\underline{\alpha }})$ we
put $\Omega ^{0}=$ $M_{k}$ and write
\begin{equation*}
\delta \varphi \left( e_{\underline{\alpha }}\right) =e_{\underline{\alpha }%
}(\varphi )
\end{equation*}%
for every matrix function $\varphi \in M_{k}(\C,u^{\underline{\alpha }}).$
As a particular case, we have
\begin{equation*}
\delta q^{\underline{\alpha }}\left( e_{\underline{\beta }}\right) =-W_{~%
\underline{\beta }\underline{\gamma }}^{\underline{\alpha }}q^{\underline{%
\gamma }}
\end{equation*}%
where indices are raised and lowered with the anholonomically deformed
metric $\rho _{\underline{\alpha }\underline{\beta }}(u^{\underline{\lambda }%
}).$ This way, we can define the set of 1--forms $\Omega ^{1}\left[ M_{k}(\C%
,u^{\underline{\alpha }})\right] $ to be the set of all elements of the form
$\varphi \delta \beta $ with $\varphi $ and $\beta $ belonging to $M_{k}(\C%
,u^{\underline{\alpha }}).$ The set of all differential forms define a
differential algebra $\Omega ^{\ast }\left[ M_{k}(\C,u^{\underline{\alpha }})%
\right] $ with the couple $\left( \Omega ^{\ast }\left[ M_{k}(\C,u^{%
\underline{\alpha }})\right] ,\delta \right) $ said to be a differential
calculus in $M_{k}(\C,u^{\underline{\alpha }})$ induced by the anholonomy of
certain exact solutions (with off--diagonal metrics and associated
N--connections) in a gravity theory.

We can also find a set of generators $e^{\underline{\alpha }}$ of $\Omega
^{1}\left[ M_{k}(\C,u^{\underline{\alpha }})\right] ,$ as a left/ right
--module completely characterized by the duality equations $e^{\underline{%
\mu }}\left( e_{\underline{\alpha }}\right) =\delta _{_{\underline{\alpha }%
}}^{\underline{\mu }}I$ and related to $\delta q^{\underline{\alpha }},$%
\begin{equation*}
\delta q^{\underline{\alpha }}=W_{~\underline{\beta }\underline{\gamma }}^{%
\underline{\alpha }}q^{\underline{\beta }}q^{\underline{\gamma }}\mbox{ and }%
e^{\underline{\mu }}=q_{\underline{\gamma }}q^{\underline{\mu }}\delta q^{%
\underline{\gamma }}.
\end{equation*}%
Similarly to the formalism presented in details in Ref. \cite{madore}, we
can elaborate a differential calculus with derivations by introducing a
linear torsionless connection%
\begin{equation*}
\mathcal{D}e^{\underline{\mu }}=-\omega _{\ \underline{\gamma }}^{\underline{%
\mu }}\otimes e^{\underline{\gamma }}
\end{equation*}%
with the coefficients $\omega _{\ \underline{\gamma }}^{\underline{\mu }}=-%
\frac{1}{2}W_{~\underline{\gamma }\underline{\beta }}^{\underline{\mu }}e^{%
\underline{\gamma }},$ resulting in the curvature 2--form%
\begin{equation*}
\mathcal{R}_{\ \underline{\gamma }}^{\underline{\mu }}=\frac{1}{8}W_{~%
\underline{\gamma }\underline{\beta }}^{\underline{\mu }}W_{~\underline{%
\alpha }\underline{\tau }}^{\underline{\beta }}e^{\underline{\alpha }}e^{%
\underline{\tau }}.
\end{equation*}
This is a surprising fact that 'commutative' curved spacetimes provided with
off--diagonal metrics and associated anhlonomic frames and N--connections
may be characterized by a noncommutative 'shadow' space with a Lie algebra
like structure induced by the frame anolonomy. We argue that such metrics
possess anholonomic noncommutative symmetries and conclude that for the
'holonomic' solutions of the Einstein equations, with vanishing $w_{~%
\underline{\alpha }\underline{\beta }}^{\underline{\gamma }},$ any
associated noncommutative geometry or $SL\left( k,\C\right) $ decouples from
the off--diagonal (anholonomic) gravitational background and transforms into
a trivial one defined by the corresponding structure constants of the chosen
Lie algebra. The anholonomic noncommutativity and the related differential
geometry are induced by the anholonomy coefficients. All such structures
reflect a specific type of symmetries of generic off--diagonal metrics and
associated frame/ N--connection structures.

Considering exact solutions of the gravitational field equations, we can
assert that we constructed a class of vacuum or nonvacuum metrics possessing
a specific noncommutative symmetry instead, for instance, of any usual
Killing symmetry. In general, we can introduce a new classification of
spacetimes following anholonomic noncommutative aglebraic properties of
metrics and vielbein structures (see Ref. \cite{vnc,vncs}). In this paper,
we analyze the simplest examples of such spacetimes.

\section{4D Static Black Ellipsoids in MAG and String Gravity}

We outline the black ellipsoid solutions \cite{velp1,velp2} and discuss
their associated anholonomic noncommutative symmetries \cite{vncs}. We note
that such solutions can be extended for the (anti) de Sitter spaces, in
gauge gravity and string gravity with effective cosmological constant \cite%
{vncfg}. In this paper, the solutions are considered for 'real'
metric--affine spaces and extended to nontrivial cosmological constant.  We
emphasize the posibility to construct solutions with locally ''anisotropic''
cosmological constants (such configurations may be also induced, for
instance, from string/ brane gravity).

\subsection{Anholonomic deformations of the Schwarzschild metric}

We consider a particular case of effectively diagonalized (\ref{dmetric4})
(and corresponding off--diagonal metric ansatz (\ref{cmetric4})) when
\begin{eqnarray}
\delta s^{2} &=&[-\left( 1-\frac{2m}{r}+\frac{\varepsilon }{r^{2}}\right)
^{-1}dr^{2}-r^{2}q(r)d\theta ^{2}  \label{sch} \\
&&-\eta _{3}\left( r,\varphi \right) r^{2}\sin ^{2}\theta d\varphi ^{2}+\eta
_{4}\left( r,\varphi \right) \left( 1-\frac{2m}{r}+\frac{\varepsilon }{r^{2}}%
\right) \delta t^{2}]  \notag
\end{eqnarray}%
where the ''polarization'' functions $\eta _{3,4}$ are decomposed on a small
parameter $\varepsilon ,0<\varepsilon \ll 1,$
\begin{eqnarray}
\eta _{3}\left( r,\varphi \right) &=&\eta _{3[0]}\left( r,\varphi \right)
+\varepsilon \lambda _{3}\left( r,\varphi \right) +\varepsilon ^{2}\gamma
_{3}\left( r,\varphi \right) +...,  \label{decom1} \\
\eta _{4}\left( r,\varphi \right) &=&1+\varepsilon \lambda _{4}\left(
r,\varphi \right) +\varepsilon ^{2}\gamma _{4}\left( r,\varphi \right) +...,
\notag
\end{eqnarray}%
and
\begin{equation*}
\delta t=dt+n_{1}\left( r,\varphi \right) dr
\end{equation*}%
for $n_{1}\sim \varepsilon ...+\varepsilon ^{2}$ terms. The functions $%
q(r),\eta _{3,4}\left( r,\varphi \right) $ and $n_{1}\left( r,\varphi
\right) $ will be found as the metric will define a solution of the
gravitational field equations generated by small deformations of the
spherical static symmetry on a small positive parameter $\varepsilon $ (in
the limits $\varepsilon \rightarrow 0$ and $q,\eta _{3,4}\rightarrow 1$ we
have just the Schwarzschild solution for a point particle of mass $m).$ The
metric (\ref{sch}) is a particular case of a class of exact solutions
constructed in \cite{v,v1,vth}. Its complexification by complex valued
N--coefficients is investigated in Ref. \cite{vncs}.

We can state a particular symmetry of the metric (\ref{sch}) by imposing a
corresponding condition of vanishing of the metric coefficient before the
term $\delta t^{2}.$ For instance, the constraints that
\begin{eqnarray}
\eta _{4}\left( r,\varphi \right) \left( 1-\frac{2m}{r}+\frac{\varepsilon }{%
r^{2}}\right)  &=&1-\frac{2m}{r}+\varepsilon \frac{\Phi _{4}}{r^{2}}%
+\varepsilon ^{2}\Theta _{4}=0,  \label{hor1} \\
\Phi _{4} &=&\lambda _{4}\left( r^{2}-2mr\right) +1  \notag \\
\Theta _{4} &=&\gamma _{4}\left( 1-\frac{2m}{r}\right) +\lambda _{4},  \notag
\end{eqnarray}%
define a rotation ellipsoid configuration if
\begin{equation*}
\lambda _{4}=\left( 1-\frac{2m}{r}\right) ^{-1}(\cos \varphi -\frac{1}{r^{2}}%
)
\end{equation*}%
and
\begin{equation*}
\gamma _{4}=-\lambda _{4}\left( 1-\frac{2m}{r}\right) ^{-1}.
\end{equation*}%
In the first order on $\varepsilon $ one obtains \ a zero value for the
coefficient before $\delta t^{2}$ if
\begin{equation}
r_{+}=\frac{2m}{1+\varepsilon \cos \varphi }=2m[1-\varepsilon \cos \varphi ],
\label{ebh}
\end{equation}%
which is the equation for a 3D ellipsoid like hypersurface with a small
eccentricity $\varepsilon .$ In general, we can consider arbitrary pairs of
functions $\lambda _{4}(r,\theta ,\varphi )$ and $\gamma _{4}(r,\theta
,\varphi )$ (for $\varphi $--anisotropies, \ $\lambda _{4}(r,\varphi )$ and $%
\gamma _{4}(r,\varphi ))$ which may be singular for some values of $r,$ or
on some hypersurvaces $r=r\left( \theta ,\varphi \right) $ ($r=r(\varphi )).$

The simplest way to define the condition of vanishing of the metric
coefficient before the value $\delta t^{2}$ is to choose $\gamma _{4}$ and $%
\lambda _{4}$ as to have $\Theta =0.$ In this case we find from \ (\ref{hor1}%
)%
\begin{equation}
r_{\pm }=m\pm m\sqrt{1-\varepsilon \frac{\Phi }{m^{2}}}=m\left[ 1\pm \left(
1-\varepsilon \frac{\Phi _{4}}{2m^{2}}\right) \right]  \label{hor1a}
\end{equation}%
where $\Phi _{4}\left( r,\varphi \right) $ is taken for $r=2m.$

For a new radial coordinate
\begin{equation}
\xi =\int dr\sqrt{|1-\frac{2m}{r}+\frac{\varepsilon }{r^{2}}|}  \label{int2}
\end{equation}%
and
\begin{equation}
h_{3}=-\eta _{3}(\xi ,\varphi )r^{2}(\xi )\sin ^{2}\theta ,\ h_{4}=1-\frac{2m%
}{r}+\varepsilon \frac{\Phi _{4}}{r^{2}},  \label{sch1q}
\end{equation}%
when $r=r\left( \xi \right) $ is inverse function after integration in (\ref%
{int2}), we write the metric (\ref{sch}) as
\begin{eqnarray}
ds^{2} &=&-d\xi ^{2}-r^{2}\left( \xi \right) q\left( \xi \right) d\theta
^{2}+h_{3}\left( \xi ,\theta ,\varphi \right) \delta \varphi
^{2}+h_{4}\left( \xi ,\theta ,\varphi \right) \delta t^{2},  \label{sch1} \\
\delta t &=&dt+n_{1}\left( \xi ,\varphi \right) d\xi ,  \notag
\end{eqnarray}%
where the coefficient $n_{1}$ is taken to solve the equation (\ref{ricci4a})
and to satisfy the (\ref{cond1}). The next step is to state the conditions
when the coefficients of metric (\ref{sch}) define solutions of the Einstein
equations. We put  $g_{1}=-1,g_{2}=-r^{2}\left( \xi \right) q\left( \xi
\right) $ and arbitrary $h_{3}(\xi ,\theta ,\varphi )$ and $h_{4}\left( \xi
,\theta \right) $ in order to find solutions the equations (\ref{ricci1a})--(%
\ref{ricci3a}). If $h_{4}$ depends on anisotropic variable $\varphi ,$ the
equation (\ref{ricci2a}) may be solved if
\begin{equation}
\sqrt{|\eta _{3}|}=\eta _{0}\left( \sqrt{|\eta _{4}|}\right) ^{\ast }
\label{conda}
\end{equation}%
for $\eta _{0}=const.$ Considering decompositions of type (\ref{decom1}) we
put $\eta _{0}=\eta /\varepsilon $ where the constant $\eta $ is taken as to
have $\sqrt{|\eta _{3}|}=1$ in the limits
\begin{equation}
\frac{\left( \sqrt{|\eta _{4}|}\right) ^{\ast }\rightarrow 0}{\varepsilon
\rightarrow 0}\rightarrow \frac{1}{\eta }=const.  \label{condb}
\end{equation}%
These conditions are satisfied if the functions $\eta _{3[0]},$ $\lambda
_{3,4}$ and $\gamma _{3,4}$ are related via relations
\begin{equation*}
\sqrt{|\eta _{3[0]}|}=\frac{\eta }{2}\lambda _{4}^{\ast },\lambda _{3}=\eta
\sqrt{|\eta _{3[0]}|}\gamma _{4}^{\ast }
\end{equation*}%
for arbitrary $\gamma _{3}\left( r,\varphi \right) .$ In this paper we
select only such solutions which satisfy the conditions (\ref{conda}) and (%
\ref{condb}).

For linear infinitezimal extensions on $\varepsilon $ of the Schwarzschild
metric, we write the solution of (\ref{ricci4a}) as
\begin{equation*}
n_{1}=\varepsilon \widehat{n}_{1}\left( \xi ,\varphi \right)
\end{equation*}%
where
\begin{eqnarray}
\widehat{n}_{1}\left( \xi ,\varphi \right) &=&n_{1[1]}\left( \xi \right)
+n_{1[2]}\left( \xi \right) \int d\varphi \ \eta _{3}\left( \xi ,\varphi
\right) /\left( \sqrt{|\eta _{4}\left( \xi ,\varphi \right) |}\right)
^{3},\eta _{4}^{\ast }\neq 0;  \label{auxf4} \\
&=&n_{1[1]}\left( \xi \right) +n_{1[2]}\left( \xi \right) \int d\varphi \
\eta _{3}\left( \xi ,\varphi \right) ,\eta _{4}^{\ast }=0;  \notag \\
&=&n_{1[1]}\left( \xi \right) +n_{1[2]}\left( \xi \right) \int d\varphi
/\left( \sqrt{|\eta _{4}\left( \xi ,\varphi \right) |}\right) ^{3},\eta
_{3}^{\ast }=0;  \notag
\end{eqnarray}%
with the functions $n_{k[1,2]}\left( \xi \right) $ to be stated by boundary
conditions.

The data
\begin{eqnarray}
g_{1} &=&-1,g_{2}=-r^{2}(\xi )q(\xi ),  \label{data} \\
h_{3} &=&-\eta _{3}(\xi ,\varphi )r^{2}(\xi )\sin ^{2}\theta ,~h_{4}=1-\frac{%
2m}{r}+\varepsilon \frac{\Phi _{4}}{r^{2}},  \notag \\
w_{1,2} &=&0,n_{1}=\varepsilon \widehat{n}_{1}\left( \xi ,\varphi \right)
,n_{2}=0,  \notag
\end{eqnarray}%
for the metric (\ref{sch}) define a class of solutions of the Einstein
equations for the canonical distinguished connection (\ref{dcon}), with
non--trivial polarization function $\eta _{3}$ and extended on parameter $%
\varepsilon $ up to the second order (the polarization functions being taken
as to make zero the second order coefficients). Such solutions are generated
by small deformations (in particular cases of rotation ellipsoid symmetry)
of the Schwarschild metric.

We can relate our solutions with some small deformations of the
Schwar\-zschild metric, as well we can satisfy the asymptotically flat
condition, if we chose such functions $n_{k[1,2]}\left( x^{i}\right) $ as $%
n_{k}\rightarrow 0$ for $\varepsilon \rightarrow 0$ and $\eta
_{3}\rightarrow 1.$ These functions have to be selected as to vanish far
away from the horizon, for instance, like $\sim 1/r^{1+\tau },\tau >0,$ for
long distances $r\rightarrow \infty .$

\subsection{Black ellipsoids and anistropic cosmological constants}

\label{belsg}

We can generalize the gravitational field equations to the gravity with
variable cosmological constants $\lambda _{\lbrack h]}\left( u^{\alpha
}\right) $ and $\lambda _{\lbrack v]}\left( u^{\alpha }\right) $ which can
be induced, for instance, from extra dimensions in string/brane gravity,
when the non-trivial components of the Einsein equations are
\begin{equation}
R_{ij}=\lambda _{\lbrack h]}\left( x^{1}\right) g_{ij}\mbox{ and }%
R_{ab}=\lambda _{\lbrack v]}(x^{k},v)g_{ab}  \label{eq17}
\end{equation}%
where Ricci tensor $R_{\mu \nu }$ with anholonomic variables has two
nontrivial components $R_{ij}$ and $R_{ab},$ and the indices take values $%
i,k=1,2$ and $a,b=3,4$ for $x^{i}=\xi $ and $y^{3}=v=\varphi $ (see
notations from the previous subsection). The equations (\ref{eq17}) contain
the equations (\ref{ricci1a}) and (\ref{ricci2a}) as particular cases when $%
\lambda _{\lbrack h]}\left( x^{1}\right) =\frac{\lambda _{\lbrack H]}^{2}}{4}
$ and $\lambda _{\lbrack v]}(x^{k},v)=\frac{\lambda _{\lbrack H]}^{2}}{4}%
+\Upsilon _{2}\left( x^{k}\right) .$

For an ansatz of type (\ref{dmetric4})
\begin{eqnarray}
\delta s^{2} &=&g_{1}(dx^{1})^{2}+g_{2}(dx^{2})^{2}+h_{3}\left(
x^{i},y^{3}\right) (\delta y^{3})^{2}+h_{4}\left( x^{i},y^{3}\right) (\delta
y^{4})^{2},  \label{ansatz18} \\
\delta y^{3} &=&dy^{3}+w_{i}\left( x^{k},y^{3}\right) dx^{i^{\prime }},\quad
\delta y^{4}=dy^{4}+n_{i}\left( x^{k},y^{3}\right) dx^{i},  \notag
\end{eqnarray}%
the Einstein equations (\ref{eq17}) are written (see \cite{vth} for details
on computation)
\begin{eqnarray}
R_{1}^{1}=R_{2}^{2}=-\frac{1}{2g_{1}g_{2}}[g_{2}^{\bullet \bullet }-\frac{%
g_{1}^{\bullet }g_{2}^{\bullet }}{2g_{1}}-\frac{(g_{2}^{\bullet })^{2}}{%
2g_{2}}+g_{1}^{^{\prime \prime }}-\frac{g_{1}^{^{\prime }}g_{2}^{^{\prime }}%
}{2g_{2}}-\frac{(g_{1}^{^{\prime }})^{2}}{2g_{1}}] &=&\lambda _{\lbrack
h]}\left( x^{k}\right) ,  \label{ricci1s} \\
R_{3}^{3}=R_{4}^{4}=-\frac{\beta }{2h_{3}h_{4}} &=&\lambda _{\lbrack
v]}(x^{k},v),  \label{ricci2s} \\
R_{3i}=-w_{i}\frac{\beta }{2h_{4}}-\frac{\alpha _{i}}{2h_{4}} &=&0,
\label{ricci3s} \\
R_{4i}=-\frac{h_{4}}{2h_{3}}\left[ n_{i}^{\ast \ast }+\gamma n_{i}^{\ast }%
\right] &=&0.  \label{ricci4s}
\end{eqnarray}%
The coefficients of equations (\ref{ricci1s}) - (\ref{ricci4s}) are given by
\begin{equation}
\alpha _{i}=\partial _{i}{h_{4}^{\ast }}-h_{4}^{\ast }\partial _{i}\ln \sqrt{%
|h_{3}h_{4}|},\qquad \beta =h_{4}^{\ast \ast }-h_{4}^{\ast }[\ln \sqrt{%
|h_{3}h_{4}|}]^{\ast },\qquad \gamma =\frac{3h_{4}^{\ast }}{2h_{4}}-\frac{%
h_{3}^{\ast }}{h_{3}}.  \label{abcs}
\end{equation}%
The various partial derivatives are denoted as $a^{\bullet }=\partial
a/\partial x^{1},a^{^{\prime }}=\partial a/\partial x^{2},a^{\ast }=\partial
a/\partial y^{3}.$ This system of equations can be solved by choosing one of
the ansatz functions (\textit{e.g.} $g_{1}\left( x^{i}\right) $ or $%
g_{2}\left( x^{i}\right) )$ and one of the ansatz functions (\textit{e.g.} $%
h_{3}\left( x^{i},y^{3}\right) $ or $h_{4}\left( x^{i},y^{3}\right) )$ to
take some arbitrary, but physically interesting form. Then, the other ansatz
functions can be analytically determined up to an integration in terms of
this choice. In this way we can generate a lot of different solutions, but
we impose the condition that the initial, arbitrary choice of the ansatz
functions is ``physically interesting'' which means that one wants to make
this original choice so that the generated final solution yield a well
behaved metric.

In this subsection, we show that the data (\ref{data}) can be extended as to
generate exact black ellipsoid solutions with nontrivial polarized
cosmological constant which can be imbedded in string theory. A complex
generalization of the solution (\ref{data}) is analyzed in Ref. \cite{vncs}
and the case locally isotropic cosmological constant was considered in Ref. %
\cite{vncfg}.

At the first \ step, we find a class of solutions with $g_{1}=-1$ and $\quad
g_{2}=g_{2}\left( \xi \right) $ solving the equation (\ref{ricci1s}), which
under such parametrizations transforms to
\begin{equation}
g_{2}^{\bullet \bullet }-\frac{(g_{2}^{\bullet })^{2}}{2g_{2}}=2g_{2}\lambda
_{\lbrack h]}\left( \xi \right) .  \label{eqaux1}
\end{equation}%
With respect to the variable $Z=(g_{2})^{2}$ this equation is written as
\begin{equation*}
Z^{\bullet \bullet }+2\lambda _{\lbrack h]}\left( \xi \right) Z=0
\end{equation*}%
which can be integrated in explicit form if $\lambda _{\lbrack h]}\left( \xi
\right) =\lambda _{\lbrack h]0}=const,$
\begin{equation*}
Z=Z_{[0]}\sin \left( \sqrt{2\lambda _{\lbrack h]0}}\xi +\xi _{\lbrack
0]}\right) ,
\end{equation*}%
for some constants $Z_{[0]}$ and $\xi _{\lbrack 0]}$ which means that
\begin{equation}
g_{2}=-Z_{[0]}^{2}\sin ^{2}\left( \sqrt{2\lambda _{\lbrack h]0}}\xi +\xi
_{\lbrack 0]}\right)  \label{aux2p}
\end{equation}%
parametrize in 'real' string gravity a class of solution of (\ref{ricci1s})
for the signature $\left( -,-,-,+\right) .$ For $\lambda _{\lbrack
h]}\rightarrow 0$ we can approximate $g_{2}=r^{2}\left( \xi \right) q\left(
\xi \right) =-\xi ^{2}$ and $Z_{[0]}^{2}=1$ which has compatibility with the
data (\ref{data}). The solution (\ref{aux2p}) with cosmological constant (of
string or non--string origin) induces oscillations in the ''horozontal''
part of the metric written with respect to N--adapted frames. If we put $%
\lambda _{\lbrack h]}\left( \xi \right) $ in (\ref{eqaux1}), we can search
the solution as $g_{2}=u^{2}$ where $u\left( \xi \right) $ solves the linear
equation%
\begin{equation*}
u^{\bullet \bullet }+\frac{\lambda _{\lbrack h]}\left( \xi \right) }{4}u=0.
\end{equation*}%
The method of integration of such equations is given in Ref. \cite{kamke}.
The explicit forms of solutions depends on function $\lambda _{\lbrack
h]}\left( \xi \right) .$ In this case we have to write \
\begin{equation}
g_{2}=r^{2}\left( \xi \right) q^{[u]}\left( \xi \right) =u^{2}\left( \xi
\right) .  \label{aux3u}
\end{equation}%
For a suitable smooth behaviour of $\lambda _{\lbrack h]}\left( \xi \right)
, $ we can generate such $u\left( \xi \right) $ and $r\left( \xi \right) $
when the $r=r\left( \xi \right) $ is the inverse function after integration
in (\ref{int2}).

The next step is to solve the equation (\ref{ricci2s}),%
\begin{equation*}
h_{4}^{\ast \ast }-h_{4}^{\ast }[\ln \sqrt{|h_{3}h_{4}|}]^{\ast }=-2\lambda
_{\lbrack v]}(x^{k},v)h_{3}h_{4}.
\end{equation*}%
For $\lambda =0$ a class of solution is given by any $\widehat{h}_{3}$ and $%
\widehat{h}_{4}$ related as
\begin{equation*}
\widehat{h}_{3}=\eta _{0}\left[ \left( \sqrt{|\hat{h}_{4}|}\right) ^{\ast }%
\right] ^{2}
\end{equation*}%
for a constant $\eta _{0}$ chosen to be negative in order to generate the
signature $\left( -,-,-,+\right) .$ For non--trivial $\lambda ,$ we may
search the solution as
\begin{equation}
h_{3}=\widehat{h}_{3}\left( \xi ,\varphi \right) ~f_{3}\left( \xi ,\varphi
\right) \mbox{ and }h_{4}=\widehat{h}_{4}\left( \xi ,\varphi \right) ,
\label{sol15}
\end{equation}%
which solves (\ref{ricci2s}) if $f_{3}=1$ for $\lambda _{\lbrack v]}=0$ and
\begin{equation*}
f_{3}=\frac{1}{4}\left[ \int \frac{\lambda _{\lbrack v]}\hat{h}_{3}\hat{h}%
_{4}}{\hat{h}_{4}^{\ast }}d\varphi \right] ^{-1}\mbox{ for }\lambda
_{\lbrack v]}\neq 0.
\end{equation*}

Now it is easy to write down the solutions of equations (\ref{ricci3s})
(being a linear equation for $w_{i})$ and (\ref{ricci4s}) (after two
integrations of $n_{i}$ on $\varphi ),$%
\begin{equation}
w_{i}=\varepsilon \widehat{w}_{i}=-\alpha _{i}/\beta ,  \label{aux3s}
\end{equation}%
were $\alpha _{i}$ and $\beta $ are computed by putting (\ref{sol15}) $\ $%
into corresponding values from (\ref{abcs}) (we chose the initial conditions
as $w_{i}\rightarrow 0$ for $\varepsilon \rightarrow 0)$ and
\begin{equation*}
n_{1}=\varepsilon \widehat{n}_{1}\left( \xi ,\varphi \right)
\end{equation*}%
where the coefficients
\begin{eqnarray}
\widehat{n}_{1}\left( \xi ,\varphi \right) &=&n_{1[1]}\left( \xi \right)
+n_{1[2]}\left( \xi \right) \int d\varphi \ ~f_{3}\left( \xi ,\varphi
\right) \eta _{3}\left( \xi ,\varphi \right) /\left( \sqrt{|\eta _{4}\left(
\xi ,\varphi \right) |}\right) ^{3},\eta _{4}^{\ast }\neq 0;  \label{aux4s}
\\
&=&n_{1[1]}\left( \xi \right) +n_{1[2]}\left( \xi \right) \int d\varphi \
~f_{3}\left( \xi ,\varphi \right) \eta _{3}\left( \xi ,\varphi \right) ,\eta
_{4}^{\ast }=0;  \notag \\
&=&n_{1[1]}\left( \xi \right) +n_{1[2]}\left( \xi \right) \int d\varphi
/\left( \sqrt{|\eta _{4}\left( \xi ,\varphi \right) |}\right) ^{3},\eta
_{3}^{\ast }=0;  \notag
\end{eqnarray}%
with the functions $n_{k[1,2]}\left( \xi \right) $ to be stated by boundary
conditions.

We conclude that the set of data $g_{1}=-1,$ with non--tivial $g_{2}\left(
\xi \right) ,h_{3},h_{4},w_{i^{\prime }}$ and $n_{1}$ stated respectively by
(\ref{aux2p}), (\ref{sol15}), (\ref{aux3s}), (\ref{aux4s}) we can define a
black ellipsoid solution with explicit dependence on polarized cosmological
''constants'' $\lambda _{\lbrack h]}\left( x^{1}\right) $ and $\lambda
_{\lbrack v]}(x^{k},v),$ i. e. a metric (\ref{ansatz18}).

Finally, we analyze the structure of noncommutative symmetries associated to
the (anti) de Sitter black ellipsoid solutions. The metric (\ref{ansatz18})
with real and/or complex coefficients defining the corresponding solutions
and its analytic extensions also do not posses Killing symmetries being
deformed by anholonomic transforms. For this solution, we can associate
certain noncommutative symmetries following the same procedure as for the
Einstein real/ complex gravity but with additional nontrivial coefficients
of anholonomy and even with nonvanishing coefficients of the nonlinear
connection curvature, $\Omega _{12}^{3}=\delta _{1}N_{2}^{3}-\delta
_{2}N_{1}^{3}.$ Taking the data (\ref{aux3s}) and (\ref{aux4s}) and formulas
(\ref{anhol}), we compute the corresponding nontrivial anholonomy
coefficients
\begin{eqnarray}
w_{\ 31}^{[N]4} &=&-w_{\ 13}^{[N]4}=\partial n_{1}\left( \xi ,\varphi
\right) /\partial \varphi =n_{1}^{\ast }\left( \xi ,\varphi \right) ,
\label{auxxx2} \\
w_{\ 12}^{[N]4} &=&-w_{\ 21}^{[N]4}=\delta _{1}(\alpha _{2}/\beta )-\delta
_{2}(\alpha _{1}/\beta )  \notag
\end{eqnarray}%
for $\delta _{1}=\partial /\partial \xi -w_{1}\partial /\partial \varphi $
and $\delta _{2}=\partial /\partial \theta -w_{2}\partial /\partial \varphi
, $ with $n_{1}$ defined by (\ref{aux4s}) and $\alpha _{1,2}$ and $\beta $
computed by using the formula (\ref{abcs}) for the solutions (\ref{sol15}).
We have a 4D exact solution with nontrivial cosmological constant. So, for $%
n+m=4,$ the condition $k^{2}-1=n+m$ can not satisfied by any integer
numbers. We may trivially extend the dimensions like $n^{\prime }=6$ and $%
m^{\prime }=m=2$ and for $k=3$ to consider the \ Lie group $SL\left( 3,\C%
\right) $ noncommutativity with corresponding values of $Q_{\underline{%
\alpha }\underline{\beta }}^{\underline{\gamma }}$\ and structure constants $%
f_{~\underline{\alpha }\underline{\beta }}^{\underline{\gamma }},$ see (\ref%
{gr1}). An extension $w_{~\alpha \beta }^{[N]\gamma }\rightarrow W_{~%
\underline{\alpha }\underline{\beta }}^{\underline{\gamma }}$ may be
performed by stating the N--deformed ''structure'' constants (\ref{anhb}), $%
W_{~\underline{\alpha }\underline{\beta }}^{\underline{\gamma }}=f_{~%
\underline{\alpha }\underline{\beta }}^{\underline{\gamma }}+w_{~\underline{%
\alpha }\underline{\beta }}^{[N]\underline{\gamma }},$ with nontrivial
values of $w_{~\underline{\alpha }\underline{\beta }}^{[N]\underline{\gamma }%
}$ given by (\ref{auxxx2}). We note that the solutions with nontrivial
cosmological constants are with induced torsion with the coefficients
computed by using formulas (\ref{torsion}) and the data (\ref{aux3u}), (\ref%
{sol15}), (\ref{aux3s}) and (\ref{aux4s}).

\subsection{ Analytic extensions of black ellipsoid metrics}

For the vacuum black ellipsoid metrics the method of analytic extension was
considered in Ref. \cite{velp1,velp2}. The coefficients of the metric (\ref%
{sch}) (equivalently (\ref{sch1})) written with respect to the \ anholonomic
frame (\ref{ddif4}) has a number of similarities with the Schwrzschild and
Reissner--N\"{o}rdstrom solutions. The cosmological ''polarized'' constants
induce some additional factors like $q^{[u]}\left( \xi \right) $ and $%
f_{3}\left( \xi ,\varphi \right) $ (see, respectively, formulas (\ref{aux3u}%
) and (\ref{sol15})) and modify the N--connection coefficients as in (\ref%
{aux3s}) and (\ref{aux4s}). For a corresponding class of smooth
polarizations, the functions $q^{[u]}$ and $f_{3}$ don not change the
singularity structure of the metric coefficients. If we identify $%
\varepsilon $ with $e^{2},$ we get a static metric with effective
''electric'' charge induced by a small, quadratic on $\varepsilon ,$
off--diagonal metric extension. The coefficients of this metric are similar
to those from the Reissner--N\"{o}rdstrom solution but additionally to the
mentioned frame anholonomy there are additional polarizations by the
functions $q^{[u]},h_{3[0]},f_{3},\eta _{3,4},w_{i}$ and $n_{1}.$ Another
very important property is that the deformed metric was stated to define a
vacuum, or with polarized cosmological constant, solution of the Einstein
equations which differs substantially from the usual Reissner--N\"{o}rdstrom
metric being an exact static solution of the Einstein--Maxwell equations.
For the limits $\varepsilon \rightarrow 0$ and $q,f_{3},h_{3[0]}\rightarrow
1 $ the metric (\ref{sch}) transforms into the usual Schwarzschild metric. A
solution with ellipsoid symmetry can be selected by a corresponding
condition of vanishing of the coefficient before the term $\delta t$ which
defines an ellipsoidal hypersurface like for the Kerr metric, but in our
case the metric is non--rotating. In general, the space may be with frame
induced torsion if we do not impose constraints on $w_{i}$ and $n_{1}$ as to
obtain vanishing nonlinear connection curvature and torsions.

The analytic extension of black ellipsoid solutions with cosmological
constant can be performed similarly both for anholonomic frames with induced
or trivial torsions. We note that the solutions in string theory may contain
a frame induced torsion with the components (\ref{torsion}) (in general, we
can consider complex coefficients, see Ref. \cite{vncs}) computed for
nontrivial $N_{i^{\prime }}^{3}=-\alpha _{i^{\prime }}/\beta $ (see (\ref%
{aux3s})) and $N_{1}^{4}=\varepsilon \widehat{n}_{1}\left( \xi ,\varphi
\right) $ (see (\ref{aux4s})). This is an explicit example illustrating that
the anholonomic frame method can be applied also for generating exact
solutions in models of gravity with nontrivial torsion. For such solutions,
we can perform corresponding analytic extensions and define Penrose diagram
formalisms if the constructions are considered with respect to N--elongated
vierbeins.

The metric (\ref{ansatz18}) has a singular behaviour for $r=r_{\pm },$ see (%
\ref{hor1a}). The aim of this subsection is to prove that this way we have
constructed a solution of the Einstein equations with polarized cosmological
constant. This solution possess an ''anisotropic'' horizon being a small
deformation on parameter $\varepsilon $ of the Schwarzschild's solution
horizon. We may analyze the anisotropic horizon's properties for some fixed
''direction'' given in a smooth vecinity of any values $\varphi =\varphi
_{0} $ and $r_{+}=r_{+}\left( \varphi _{0}\right) .$ $\ $The final
conclusions will be some general ones for arbitrary $\varphi $ when the
explicit values of coefficients will have a parametric dependence on angular
coordinate $\varphi .$ The metrics (\ref{sch}), or (\ref{sch1}), and (\ref%
{ansatz18}) are regular in the regions I ($\infty >r>r_{+}^{\Phi }),$ II ($%
r_{+}^{\Phi }>r>r_{-}^{\Phi })$ and III$\;(r_{-}^{\Phi }>r>0).$ As in the
Schwarzschild, Reissner--N\"{o}rdstrom and Kerr cases these singularities
can be removed by introducing suitable coordinates and extending the
manifold to obtain a maximal analytic extension \cite{gb,carter}. We have
similar regions as in the Reissner--N\"{o}rdstrom space--time, but with just
only one possibility $\varepsilon <1$ instead of three relations for static
electro--vacuum cases ($e^{2}<m^{2},e^{2}=m^{2},e^{2}>m^{2};$ where $e$ and $%
m$ are correspondingly the electric charge and mass of the point particle in
the Reissner--N\"{o}rdstrom metric). So, we may consider the usual Penrose's
diagrams as for a particular case of the Reissner--N\"{o}rdstrom space--time
but keeping in mind that such diagrams and horizons have an additional
polarizations and parametrization on an angular coordinate.

We can proceed in steps analogous to those in the Schwarzschild case (see
details, for instance, in Ref. \cite{haw})) in order to construct the
maximally extended manifold. The first step is to introduce a new coordinate
\begin{equation*}
r^{\Vert }=\int dr\left( 1-\frac{2m}{r}+\frac{\varepsilon }{r^{2}}\right)
^{-1}
\end{equation*}%
for $r>r_{+}^{1}$ and find explicitly the coodinate
\begin{equation}
r^{\Vert }=r+\frac{(r_{+}^{1})^{2}}{r_{+}^{1}-r_{-}^{1}}\ln (r-r_{+}^{1})-%
\frac{(r_{-}^{1})^{2}}{r_{+}^{1}-r_{-}^{1}}\ln (r-r_{-}^{1}),  \label{r1}
\end{equation}%
where $r_{\pm }^{1}=r_{\pm }^{\Phi }$ with $\Phi =1.$ If $r$ is expressed as
a function on $\xi ,$ than $r^{\Vert }$ can be also expressed as a function
on $\xi $ depending additionally on some parameters.

Defining the advanced and retarded coordinates, $v=t+r^{\Vert }$ and $%
w=t-r^{\Vert },$ with corresponding elongated differentials
\begin{equation*}
\delta v=\delta t+dr^{\Vert }\mbox{ and }\delta w=\delta t-dr^{\Vert }
\end{equation*}%
the metric (\ref{sch1}) takes the form%
\begin{equation}
\delta s^{2}=-r^{2}(\xi )q^{[u]}(\xi )d\theta ^{2}-\eta _{3}(\xi ,\varphi
_{0})f_{3}(\xi ,\varphi _{0})r^{2}(\xi )\sin ^{2}\theta \delta \varphi
^{2}+(1-\frac{2m}{r(\xi )}+\varepsilon \frac{\Phi _{4}(r,\varphi _{0})}{%
r^{2}(\xi )})\delta v\delta w,  \label{met5a}
\end{equation}%
where (in general, in non--explicit form) $r(\xi )$ is a function of type $%
r(\xi )=r\left( r^{\Vert }\right) =$ $r\left( v,w\right) .$ Introducing new
coordinates $(v^{\prime \prime },w^{\prime \prime })$ by%
\begin{equation*}
v^{\prime \prime }=\arctan \left[ \exp \left( \frac{r_{+}^{1}-r_{-}^{1}}{%
4(r_{+}^{1})^{2}}v\right) \right] ,w^{\prime \prime }=\arctan \left[ -\exp
\left( \frac{-r_{+}^{1}+r_{-}^{1}}{4(r_{+}^{1})^{2}}w\right) \right]
\end{equation*}%
Defining $r$ by
\begin{equation*}
\tan v^{\prime \prime }\tan w^{\prime \prime }=-\exp \left[ \frac{%
r_{+}^{1}-r_{-}^{1}}{2(r_{+}^{1})^{2}}r\right] \sqrt{\frac{r-r_{+}^{1}}{%
(r-r_{-}^{1})^{\chi }}},\chi =\left( \frac{r_{+}^{1}}{r_{-}^{1}}\right) ^{2}
\end{equation*}
and multiplying (\ref{met5a}) on the conformal factor we obtain
\begin{eqnarray}
\delta s^{2} &=&-r^{2}q^{[u]}(r)d\theta ^{2}-\eta _{3}(r,\varphi
_{0})f_{3}(r,\varphi _{0})r^{2}\sin ^{2}\theta \delta \varphi ^{2}
\label{el2b} \\
&&+64\frac{(r_{+}^{1})^{4}}{(r_{+}^{1}-r_{-}^{1})^{2}}(1-\frac{2m}{r(\xi )}%
+\varepsilon \frac{\Phi _{4}(r,\varphi _{0})}{r^{2}(\xi )})\delta v^{\prime
\prime }\delta w^{\prime \prime },  \notag
\end{eqnarray}%
As particular cases, we may chose $\eta _{3}\left( r,\varphi \right) $ as
the condition of vanishing of the metric coefficient before $\delta
v^{\prime \prime }\delta w^{\prime \prime }$ will describe a horizon
parametrized by a resolution ellipsoid hypersurface. We emphasize that
quadratic elements (\ref{met5a}) and (\ref{el2b}) have respective
coefficients as the metrics investigated in\ Refs. \cite{velp1,velp2} but
the polarized cosmological constants introduce not only additional
polarizing factors $q^{[u]}(r)$ and $f_{3}(r,\varphi _{0})$ but also
elongate the anholonomic frames in a different manner.

The maximal \ extension of the Schwarzschild metric deformed by a small
parameter $\varepsilon $ (for ellipsoid configurations treated as the
eccentricity), i. e. \ the extension of the metric (\ref{ansatz18}), is
defined by taking (\ref{el2b}) as the metric on the maximal manifold on
which this metric is of smoothly class $C^{2}.$ The Penrose diagram of this
static but locally anisotropic space--time, for any fixed angular value $%
\varphi _{0}$ is similar to the Reissner--Nordstrom solution, for the case $%
e^{2}\rightarrow \varepsilon $ and $e^{2}<m^{2}$(see, for instance, Ref. %
\cite{haw})). There are an infinite number of asymptotically flat regions of
type I, connected by intermediate regions II and III, where there is still
an irremovable singularity at $r=0$ for every region III. We may travel from
a region I to another ones by passing through the 'wormholes' made by
anisotropic deformations (ellipsoid off--diagonality of metrics, or
anholonomy) like in the Reissner--Nordstrom universe because $\sqrt{%
\varepsilon }$ may model an effective electric charge. One can not turn back
in a such travel. Of course, this interpretation holds true only for a
corresponding smoothly class of polarization functions. For instance, if the
cosmological constant is periodically polarized from a string model, see the
formula (\ref{eqaux1}), one could be additional resonances, aperiodicity and
singularities.

It should be noted that the metric (\ref{el2b}) can be analytic every were
except at $r=r_{-}^{1}.$ We may eliminate this coordinate degeneration by
introducing another new coordinates%
\begin{equation*}
v^{\prime \prime \prime }=\arctan \left[ \exp \left( \frac{%
r_{+}^{1}-r_{-}^{1}}{2n_{0}(r_{+}^{1})^{2}}v\right) \right] ,w^{\prime
\prime \prime }=\arctan \left[ -\exp \left( \frac{-r_{+}^{1}+r_{-}^{1}}{%
2n_{0}(r_{+}^{1})^{2}}w\right) \right] ,
\end{equation*}%
where the integer $n_{0}\geq (r_{+}^{1})^{2}/(r_{-}^{1})^{2}.$ In \ these
coordinates, the metric is analytic every were except at $r=r_{+}^{1}$ where
it is degenerate.$\,$\ This way the space--time manifold can be covered by
an analytic atlas by using coordinate carts defined by $(v^{\prime \prime
},w^{\prime \prime },\theta ,\varphi )$ and $(v^{\prime \prime \prime
},w^{\prime \prime \prime },\theta ,\varphi ).$ Finally, we note that the
analytic extensions of the deformed metrics were performed with respect to
anholonomc frames which distinguish such constructions from those dealing
only with holonomic coordinates, like for the usual Reissner--N\"{o}rdstrom
and Kerr metrics. We stated the conditions when on 'radial' like coordinates
we preserve the main properties of the well know black hole solutions but in
our case the metrics are generic off--diagonal and with vacuum gravitational
polarizations.

\subsection{Geodesics on static polarized ellipsoid backgrounds}

We analyze the geodesic congruence of the metric (\ref{ansatz18}) with the
data (\ref{data}) modified by polarized cosmological constant, for
simplicity, being linear on $\varepsilon ,$by introducing the effective
Lagrangian (for instance, like in Ref. \cite{chan})%
\begin{eqnarray}
2L &=&g_{\alpha \beta }\frac{\delta u^{\alpha }}{ds}\frac{\delta u^{\beta }}{%
ds}=-\left( 1-\frac{2m}{r}+\frac{\varepsilon }{r^{2}}\right) ^{-1}\left(
\frac{dr}{ds}\right) ^{2}-r^{2}q^{[u]}(r)\left( \frac{d\theta }{ds}\right)
^{2}  \label{lagrb} \\
&&-\eta _{3}(r,\varphi )f_{3}(r,\varphi )r^{2}\sin ^{2}\theta \left( \frac{%
d\varphi }{ds}\right) ^{2}+\left( 1-\frac{2m}{r}+\frac{\varepsilon \Phi _{4}%
}{r^{2}}\right) \left( \frac{dt}{ds}+\varepsilon \widehat{n}_{1}\frac{dr}{ds}%
\right) ^{2},  \notag
\end{eqnarray}%
for $r=r(\xi ).$

The corresponding Euler--Lagrange equations,
\begin{equation*}
\frac{d}{ds}\frac{\partial L}{\partial \frac{\delta u^{\alpha }}{ds}}-\frac{%
\partial L}{\partial u^{\alpha }}=0
\end{equation*}%
are%
\begin{eqnarray}
&&\frac{d}{ds}\left[ -r^{2}q^{[u]}(r)\frac{d\theta }{ds}\right] =-\eta
_{3}f_{3}r^{2}\sin \theta \cos \theta \left( \frac{d\varphi }{ds}\right)
^{2},  \label{lag2b} \\
&&\frac{d}{ds}\left[ -\eta _{3}f_{3}r^{2}\frac{d\varphi }{ds}\right] =-(\eta
_{3}f_{3})^{\ast }\frac{r^{2}}{2}\sin ^{2}\theta \left( \frac{d\varphi }{ds}%
\right) ^{2}+\frac{\varepsilon }{2}\left( 1-\frac{2m}{r}\right) \left[ \frac{%
\Phi _{4}^{\ast }}{r^{2}}\left( \frac{dt}{ds}\right) ^{2}+\widehat{n}%
_{1}^{\ast }\frac{dt}{ds}\frac{d\xi }{ds}\right]  \notag
\end{eqnarray}%
and%
\begin{equation}
\frac{d}{ds}\left[ (1-\frac{2m}{r}+\frac{\varepsilon \Phi _{4}}{r^{2}}%
)\left( \frac{dt}{ds}+\varepsilon \widehat{n}_{1}\frac{d\xi }{ds}\right) %
\right] =0,  \label{lag3}
\end{equation}%
where, for instance, $\Phi _{4}^{\ast }=\partial $ $\Phi _{4}/\partial
\varphi $ we have omitted the variations for $d\xi /ds$ which may be found
from (\ref{lagrb}). The sistem of equations (\ref{lagrb})--(\ref{lag3})
transform into the usual system of geodesic equations for the Schwarzschild
space--time if $\varepsilon \rightarrow 0$ and $q^{[u]},\eta
_{3},f_{3}\rightarrow 1$ which can be solved exactly \cite{chan}. For
nontrivial values of the parameter $\varepsilon $ and polarizations $\eta
_{3},f_{3}$ even to obtain some decompositions of solutions on $\varepsilon $
for arbitrary $\eta _{3}$ and $n_{1[1,2]},$ see (\ref{auxf4}), is a
cumbersome task. In spite of the fact that with respect to anholonomic
frames the metrics (\ref{sch1}) and/or (\ref{ansatz18}) has their
coefficients \ being very similar to the Reissner--Nordstom solution. The
geodesic behaviour, in our anisotropic cases, is more sophisticate because
of anholonomy, polarization of constants and coefficients and ''elongation''
of partial derivatives. For instance, the equations (\ref{lag2b}) \ state
that there is not any angular on $\varphi ,$ conservation law if $(\eta
_{3}f_{3})^{\ast }\neq 0,$ even for $\varepsilon \rightarrow 0$ (which holds
both for the Schwarzschild and Reissner--Nordstom metrics). One follows from
the equation (\ref{lag3}) the existence of an energy like integral of
motion, $E=E_{0}+$ $\varepsilon E_{1},$ with%
\begin{eqnarray*}
E_{0} &=&\left( 1-\frac{2m}{r}\right) \frac{dt}{ds} \\
E_{1} &=&\frac{\Phi _{4}}{r^{2}}\frac{dt}{ds}+\left( 1-\frac{2m}{r}\right)
\widehat{n}_{1}\frac{d\xi }{ds}.
\end{eqnarray*}

The introduced anisotropic deformations of congruences of Schwarzschild's
space--time geodesics mantain the known behaviour in the vecinity of the
horizon hypersurface defined by the condition of vanishing of the
coefficient $\left( 1-2m/r+\varepsilon \Phi _{4}/r^{2}\right) $ in (\ref%
{el2b}). The simplest way to prove this is to consider radial null geodesics
in the ''equatorial plane'', which satisfy the condition (\ref{lagrb}) with $%
\theta =\pi /2,d\theta /ds=0,d^{2}\theta /ds^{2}=0$ and $d\varphi /ds=0,$
from which follows that%
\begin{equation*}
\frac{dr}{dt}=\pm \left( 1-\frac{2m}{r}+\frac{\varepsilon _{0}}{r^{2}}%
\right) \left[ 1+\varepsilon \widehat{n}_{1}d\varphi \right] .
\end{equation*}%
The integral of this equation, for every fixed value $\varphi =\varphi _{0}$
is
\begin{equation*}
t=\pm r^{\Vert }+\varepsilon \int \left[ \frac{\Phi _{4}(r,\varphi _{0})-1}{%
2\left( r^{2}-2mr\right) ^{2}}-\widehat{n}_{1}(r,\varphi _{0})\right] dr
\end{equation*}%
where the coordinate $r^{\Vert }$ is defined in equation (\ref{r1}). In this
formula the term proportional to $\varepsilon $ can have non--singular
behaviour for a corresponding class of polarizations $\lambda _{4},$ see the
formulas (\ref{hor1}). Even the explicit form of the integral depends on the
type of polarizations $\eta _{3}(r,\varphi _{0}),$ $f_{3}(r,\varphi _{0})$
and values $n_{1[1,2]}(r),$ which results in some small deviations of the
null--geodesics, we may conclude that for an in--going null--ray the
coordinate time $t$ increases from $-\infty $ to $+\infty $ as $r$ decreases
from $+\infty $ to $r_{+}^{1},$ decreases from $+\infty $ to $-\infty $ as $%
r $ further decreases from $r_{+}^{1}$ to $r_{-}^{1},$ and increases again
from $-\infty $ to a finite limit as $r$ decreses from $r_{-}^{1}$ to zero.
We have a similar behaviour as for the Reissner--Nordstrom solution but with
some additional anisotropic contributions being proportional to $\varepsilon
.$ Here we also note that as $dt/ds$ tends to $+\infty $ for $r\rightarrow
r_{+}^{1}+0$ and to $-\infty $ as $r_{-}+0,$ any radiation received from
infinity appear to be infinitely red--shifted at the crossing of the event
horizon and infinitely blue--shifted at the crossing of the Cauchy horizon.

The mentioned properties of null--geodesics allow us to conclude that the
metric (\ref{sch}) (equivalently, (\ref{sch1})) with the data (\ref{data})
and their maximal analytic extension (\ref{el2b}) really define a black hole
static solution which is obtained by anisotropic small deformations on $%
\varepsilon $ and renormalization by $\eta _{3}f_{3}$ of the Schwarzchild
solution (for a corresponding type of deformations the horizon of such black
holes is defined by ellipsoid hypersurfaces). We call such objects as black
ellipsoids, or black rotoids. They exists in the framework of general
relativity as certain solutions of the Einstein equations defined by static
generic off--diagonal metrics and associated anholonomic frames or can be
induced by polarized cosmological constants. This property disinguishes them
from similar configurations of Reissner--Norstrom type (which are static
electrovacuum solutions of the Einstein--Maxwell equations) and of Kerr type
rotating configurations, with ellipsoid horizon, also defined by
off--diagonal vacuum metrics (here we emphasized that the spherical
coordinate system is associated to a holonomic frame which is a trivial case
of anholonomic bases). By introducing the polarized cosmological constants,
the anholonomic character of N--adapted frames allow to construct solutions
being very different from the black hole solutions in (anti) de Sitter
spacetimes. We selected here a class of solutions where cosmological factors
correspond to some additional polarizations but do not change the
singularity structure of black ellipsoid solutions.

The metric (\ref{sch}) and its analytic extensions do not posses Killing
symmetries being deformed by anholonomic transforms. Nevertheless, we can
associate to such solutions certain noncommutative symmetries \cite{vncs}.
Taking the data (\ref{data}) and formulas (\ref{anh}), we compute the
corresponding nontrivial anholonomy coefficients
\begin{equation}
w_{\ 42}^{[N]5}=-w_{\ 24}^{[N]5}=\partial n_{2}\left( \xi ,\varphi \right)
/\partial \varphi =n_{2}^{\ast }\left( \xi ,\varphi \right)  \label{auxxx1}
\end{equation}%
with $n_{2}$ defined by (\ref{data}). Our solutions are for 4D
configuration. So for $n+m=4,$ the condition $k^{2}-1=n+m$ can not satisfied
in integer numbers. We may trivially extend the dimensions like $n^{\prime
}=6$ and $m^{\prime }=m=2 $ and for $k=3$ to consider the \ Lie group $%
SL\left( 3,\C\right) $ noncommutativity with corresponding values of $Q_{%
\underline{\alpha }\underline{\beta }}^{\underline{\gamma }}$\ and structure
constants $f_{~\underline{\alpha }\underline{\beta }}^{\underline{\gamma }},$
see (\ref{gr1}). An extension $w_{~\alpha \beta }^{[N]\gamma }\rightarrow
W_{~\underline{\alpha }\underline{\beta }}^{\underline{\gamma }}$ may be
performed by stating the N--deformed ''structure'' constants (\ref{anhb}), $%
W_{~\underline{\alpha }\underline{\beta }}^{\underline{\gamma }}=f_{~%
\underline{\alpha }\underline{\beta }}^{\underline{\gamma }}+w_{~\underline{%
\alpha }\underline{\beta }}^{[N]\underline{\gamma }},$ with only two
nontrivial values of $w_{~\underline{\alpha }\underline{\beta }}^{[N]%
\underline{\gamma }}$ given by (\ref{auxxx1}). In a similar manner we can
compute the anholonomy coefficients for the black ellipsoid metric with
cosmological constant contributions (\ref{ansatz18}).

\section{Perturbations of Anisotropic Black Holes}

The stablility of black ellipsoids was proven in Ref. \cite{vels}. A similar
proof may hold true for a class of metrics with anholonomic noncommutative
symmetry and possible complexification of some off--diagonal metric and
tetradics coefficients \cite{vncs}. In this section we reconsider the
perturbation formalism and stability proofs for rotoid metrics defined by
polarized cosmological constants.

\subsection{Metrics describing anisotropic perturbations}

We consider a four dimensional pseudo--Riemannian quadratic linear element
\begin{eqnarray}
ds^{2} &=&\Omega (r,\varphi )\left[ -\left( 1-\frac{2m}{r}+\frac{\varepsilon
}{r^{2}}\right) ^{-1}dr^{2}-r^{2}q^{[v]}(r)d\theta ^{2}-\eta
_{3}^{[v]}(r,\theta ,\varphi )r^{2}\sin ^{2}\theta \delta \varphi ^{2}\right]
\notag \\
&&+\left[ 1-\frac{2m}{r}+\frac{\varepsilon }{r^{2}}\eta (r,\varphi )\right]
\delta t^{2},  \label{metric1p} \\
\eta _{3}^{[v]}(r,\theta ,\varphi ) &=&\eta _{3}(r,\theta ,\varphi
)f_{3}(r,\theta ,\varphi )  \notag
\end{eqnarray}%
with
\begin{equation*}
\delta \varphi =d\varphi +\varepsilon w_{1}(r,\varphi )dr,\mbox{
and }\delta t=dt+\varepsilon n_{1}(r,\varphi )dr,
\end{equation*}%
where the local coordinates are denoted $u=\{u^{\alpha }=\left( r,\theta
,\varphi ,t\right) \}$ (the Greek indices $\alpha ,\beta ,...$ will run the
values 1,2,3,4), $\varepsilon $ is a small parameter satisfying the
conditions $0\leq \varepsilon \ll 1$ (for instance, an eccentricity for some
ellipsoid deformations of the spherical symmetry) and the functions $\Omega
(r,\varphi ),q(r),$ $\eta _{3}(r,\theta ,\varphi )$ and $\eta (\theta
,\varphi )$ are of necessary smooth class. The metric (\ref{metric1p}) is
static, off--diagonal and transforms into the usual Schwarzschild solution
if $\varepsilon \rightarrow 0$ and $\Omega ,q^{[v]},\eta
_{3}^{[v]}\rightarrow 1.$ For vanishing cosmological constants, it describes
at least two classes of static black hole solutions generated as small
anhlonomic deformations of the Schwarzschild solution \cite%
{v,v1,velp1,velp2,vncfg,vels} but models also nontrivial vacuum polarized
cosmological constants.

We can apply the perturbation theory for the metric (\ref{metric1p}) (not
paying a special attention to some particular parametrization of
coefficients for one or another class of anisotropic black hole solutions)
and analyze its stability by using the results of Ref. \cite{chan} for a
fixed anisotropic direction, i. e. by imposing certain anholonomic frame
constraints for an angle $\varphi =\varphi _{0}$ but considering possible
perturbations depending on three variables $(u^{1}=x^{1}=r,u^{2}=x^{2}=%
\theta ,$ $u^{4}=t).$ We suppose that if we prove that there is a stability
on perturbations for a value $\varphi _{0},$ we can analyze in a similar
manner another values of $\varphi .$ A more general perturbative theory with
variable anisotropy on coordinate $\varphi ,$ i. e. with dynamical
anholonomic constraints, connects the approach with a two dimensional
inverse problem which makes the analysis more sophisticate. There have been
not elaborated such analytic methods in the theory of black holes.

It should be noted that in a study of perturbations of any spherically
symmetric system \ and, for instance, of small ellipsoid deformations,
without any loss of generality, we can restrict our considerations to
axisymmetric modes of perturbations. Non--axisymmetric modes of
perturbations with an $e^{in\varphi }$ dependence on the azimutal angle $%
\varphi $ $\ (n$ being an integer number) can be deduced from modes of
axisymmetric perturbations with $n=0$ by suitable rotations since there are
not preferred axes in a spherically symmetric background. The ellipsoid like
deformations may be included into the formalism as some low frequency and
constrained petrurbations.

We develope the black hole perturbation and stability theory as to include
into consideration off--diagonal metrics with the coefficients polarized by
cosmological constants. This is the main difference comparing to the paper %
\cite{vels}. For simplicity, in this section, we restrict our study only to
fixed values of the coordinate $\varphi $ assuming that anholonomic
deformations are proportional to a small parameter $\varepsilon ;$ we shall
investigate the stability of solutions only by applying the one dimensional
inverse methods.

We state a quadratic metric element
\begin{eqnarray}
ds^{2} &=&-e^{2\mu _{1}}(du^{1})^{2}-e^{2\mu _{2}}(du^{2})^{2}-e^{2\mu
_{3}}(\delta u^{3})^{2}+e^{2\mu _{4}}(\delta u^{4})^{2},  \notag \\
\delta u^{3} &=&d\varphi -q_{1}dx^{1}-q_{2}dx^{2}-\omega dt,  \label{metric2}
\\
\delta u^{4} &=&dt+n_{1}dr  \notag
\end{eqnarray}%
where
\begin{eqnarray}
\mu _{\alpha }(x^{k},t) &=&\mu _{\alpha }^{(\varepsilon )}(x^{k},\varphi
_{0})+\delta \mu _{\alpha }^{(\varsigma )}(x^{k},t),  \label{coef1} \\
q_{i}(x^{k},t) &=&q_{i}^{(\varepsilon )}(r,\varphi _{0})+\delta
q_{i}^{(\varsigma )}(x^{k},t),\ \omega (x^{k},t)=0+\delta \omega
^{(\varsigma )}(x^{k},t)  \notag
\end{eqnarray}%
with
\begin{eqnarray}
e^{2\mu _{1}^{(\varepsilon )}} &=&\Omega (r,\varphi _{0})(1-\frac{2m}{r}+%
\frac{\varepsilon }{r^{2}})^{-1},\ e^{2\mu _{2}^{(\varepsilon )}}=\Omega
(r,\varphi _{0})q^{[v]}(r)r^{2},  \label{coef2} \\
e^{2\mu _{3}^{(\varepsilon )}} &=&\Omega (r,\varphi _{0})r^{2}\sin
^{2}\theta \eta _{3}^{[v]}(r,\varphi _{0}),\ e^{2\mu _{4}^{(\varepsilon
)}}=1-\frac{2m}{r}+\frac{\varepsilon }{r^{2}}\eta (r,\varphi _{0}),  \notag
\end{eqnarray}%
and some non--trivial values for $q_{i}^{(\varepsilon )}$ and $\varepsilon
n_{i},$%
\begin{eqnarray*}
q_{i}^{(\varepsilon )} &=&\varepsilon w_{i}(r,\varphi _{0}), \\
n_{1} &=&\varepsilon \left( n_{1[1]}(r)+n_{1[2]}(r)\int_{0}^{\varphi
_{0}}\eta _{3}(r,\varphi )d\varphi \right) .
\end{eqnarray*}%
We have to distinguish two types of small deformations from the spherical
symmetry. The first type of deformations, labeled with the index $%
(\varepsilon )$ are generated by some $\varepsilon $--terms which define a
fixed ellipsoid like configuration and the second type ones, labeled with
the index $(\varsigma ),$ are some small linear fluctuations of the metric
coefficients

The general formluas for the Ricci and Einstein tensors for metric elements
of class (\ref{metric2}) with $w_{i},n_{1}=0$ are given in \cite{chan}. We
compute similar values with respect to anholnomic frames, when, for a
conventional splitting $u^{\alpha }=(x^{i},y^{a}),$ the coordinates $x^{i}$
and $y^{a}$ are treated respectively as holonomic and anholonomic ones. In
this case the partial derivatives $\partial /\partial x^{i}$ must be changed
into certain 'elongated' ones
\begin{eqnarray*}
\frac{\partial }{\partial x^{1}} &\rightarrow &\frac{\delta }{\partial x^{1}}%
=\frac{\partial }{\partial x^{1}}-w_{1}\frac{\partial }{\partial \varphi }%
-n_{1}\frac{\partial }{\partial t}, \\
\frac{\partial }{\partial x^{2}} &\rightarrow &\frac{\delta }{\partial x^{2}}%
=\frac{\partial }{\partial x^{2}}-w_{2}\frac{\partial }{\partial \varphi },
\end{eqnarray*}%
see details in Refs \cite{vth,vsbd,velp1,velp2}. In the ansatz (\ref{metric2}%
), the anholonomic contributions of $w_{i}$ are included in the coefficients
$q_{i}(x^{k},t).$ For convenience, we give present bellow the necessary
formulas for $R_{\alpha \beta }$ (the Ricci tensor) and $G_{\alpha \beta }$
(the Einstein tensor) computed for the ansatz (\ref{metric2}) with three
holonomic coordinates $\left( r,\theta ,\varphi \right) $ and on anholonomic
coodinate $t$ (in our case, being time like), with the partial derivative
operators
\begin{equation*}
\partial _{1}\rightarrow \delta _{1}=\frac{\partial }{\partial r}w_{1}\frac{%
\partial }{\partial \varphi }-n_{1}\frac{\partial }{\partial t},\delta _{2}=%
\frac{\partial }{\partial \theta }-w_{2}\frac{\partial }{\partial \varphi }%
,\partial _{3}=\frac{\partial }{\partial \varphi },
\end{equation*}%
and for a fixed value $\varphi _{0}.$

A general perturbation of an anisotropic black--hole described by a
quadratic line element (\ref{metric2}) \ results in some small quantities of
the first order $\omega $ and $q_{i},$ inducing a dragging of frames and
imparting rotations, and in some functions $\mu _{\alpha }$ with small
increments $\delta \mu _{\alpha },$ which do not impart rotations. Some
coefficients contained in such values are proportional to $\varepsilon ,$
another ones are considered only as small quantities. The perturbations of
metric are of two generic types: axial and polar one. We shall ivestigate
them separately in the next two subsection after we shall have computed the
coefficients of the Ricci tensor.

We compute the coefficients of the the Ricci tensor as
\begin{equation*}
R_{\beta \gamma \alpha }^{\alpha }=R_{\beta \gamma }
\end{equation*}%
and of the Einstein tensor as%
\begin{equation*}
G_{\beta \gamma }=R_{\beta \gamma }-\frac{1}{2}g_{\beta \gamma }R
\end{equation*}%
for $R=g^{\beta \gamma }R_{\beta \gamma }.$ Straightforward computations for
the quadratic line element (\ref{metric2}) give%
\begin{eqnarray}
R_{11} &=&-e^{-2\mu _{1}}[\delta _{11}^{2}(\mu _{3}+\mu _{4}+\mu
_{2})+\delta _{1}\mu _{3}\delta _{1}(\mu _{3}-\mu _{1})+\delta _{1}\mu
_{2}\delta _{1}(\mu _{2}-\mu _{1})+  \label{riccip} \\
&&\delta _{1}\mu _{4}\delta _{1}(\mu _{4}-\mu _{1})]-e^{-2\mu _{2}}[\delta
_{22}^{2}\mu _{1}+\delta _{2}\mu _{1}\delta _{2}(\mu _{3}+\mu _{4}+\mu
_{1}-\mu _{2})]+  \notag \\
&&e^{-2\mu _{4}}[\partial _{44}^{2}\mu _{1}+\partial _{4}\mu _{1}\partial
_{4}(\mu _{3}-\mu _{4}+\mu _{1}+\mu _{2})]-\frac{1}{2}e^{2(\mu _{3}-\mu
_{1})}[e^{-2\mu _{2}}Q_{12}^{2}+e^{-2\mu _{4}}Q_{14}^{2}],  \notag
\end{eqnarray}%
\begin{eqnarray*}
R_{12} &=&-e^{-\mu _{1}-\mu _{2}}[\delta _{2}\delta _{1}(\mu _{3}+\mu
_{2})-\delta _{2}\mu _{1}\delta _{1}(\mu _{3}+\mu _{1})-\delta _{1}\mu
_{2}\partial _{4}(\mu _{3}+\mu _{1}) \\
&&+\delta _{1}\mu _{3}\delta _{2}\mu _{3}+\delta _{1}\mu _{4}\delta _{2}\mu
_{4}]+\frac{1}{2}e^{2\mu _{3}-2\mu _{4}-\mu _{1}-\mu _{2}}Q_{14}Q_{24},
\end{eqnarray*}%
\begin{equation*}
R_{31}=-\frac{1}{2}e^{2\mu _{3}-\mu _{4}-\mu _{2}}[\delta _{2}(e^{3\mu
_{3}+\mu _{4}-\mu _{1}-\mu _{2}}Q_{21})+\partial _{4}(e^{3\mu _{3}-\mu
_{4}+\mu _{2}-\mu _{1}}Q_{41})],
\end{equation*}%
\begin{eqnarray*}
R_{33} &=&-e^{-2\mu _{1}}[\delta _{11}^{2}\mu _{3}+\delta _{1}\mu _{3}\delta
_{1}(\mu _{3}+\mu _{4}+\mu _{2}-\mu _{1})]- \\
&&e^{-2\mu _{2}}[\delta _{22}^{2}\mu _{3}+\partial _{2}\mu _{3}\partial
_{2}(\mu _{3}+\mu _{4}-\mu _{2}+\mu _{1})]+\frac{1}{2}e^{2(\mu _{3}-\mu
_{1}-\mu _{2})}Q_{12}^{2}+ \\
&&e^{-2\mu _{4}}[\partial _{44}^{2}\mu _{3}+\partial _{4}\mu _{3}\partial
_{4}(\mu _{3}-\mu _{4}+\mu _{2}+\mu _{1})]-\frac{1}{2}e^{2(\mu _{3}-\mu
_{4})}[e^{-2\mu _{2}}Q_{24}^{2}+e^{-2\mu _{1}}Q_{14}^{2}],
\end{eqnarray*}%
\begin{eqnarray*}
R_{41} &=&-e^{-\mu _{1}-\mu _{4}}[\partial _{4}\delta _{1}(\mu _{3}+\mu
_{2})+\delta _{1}\mu _{3}\partial _{4}(\mu _{3}-\mu _{1})+\delta _{1}\mu
_{2}\partial _{4}(\mu _{2}-\mu _{1}) \\
&&-\delta _{1}\mu _{4}\partial _{4}(\mu _{3}+\mu _{2})]+\frac{1}{2}e^{2\mu
_{3}-\mu _{4}-\mu _{1}-2\mu _{2}}Q_{12}Q_{34},
\end{eqnarray*}%
\begin{equation*}
R_{43}=-\frac{1}{2}e^{2\mu _{3}-\mu _{1}-\mu _{2}}[\delta _{1}(e^{3\mu
_{3}-\mu _{4}-\mu _{1}+\mu _{2}}Q_{14})+\delta _{2}(e^{3\mu _{3}-\mu
_{4}+\mu _{1}-\mu _{2}}Q_{24})],
\end{equation*}%
\begin{eqnarray*}
R_{44} &=&-e^{-2\mu _{4}}[\partial _{44}^{2}(\mu _{1}+\mu _{2}+\mu
_{3})+\partial _{4}\mu _{3}\partial _{4}(\mu _{3}-\mu _{4})+\partial _{4}\mu
_{1}\partial _{4}(\mu _{1}-\mu _{4})+ \\
&&\partial _{4}\mu _{2}\partial _{4}(\mu _{2}-\mu _{4})]+e^{-2\mu
_{1}}[\delta _{11}^{2}\mu _{4}+\delta _{1}\mu _{4}\delta _{1}(\mu _{3}+\mu
_{4}-\mu _{1}+\mu _{2})]+ \\
&&e^{-2\mu _{2}}[\delta _{22}^{2}\mu _{4}+\delta _{2}\mu _{4}\delta _{2}(\mu
_{3}+\mu _{4}-\mu _{1}+\mu _{2})]-\frac{1}{2}e^{2(\mu _{3}-\mu
_{4})}[e^{-2\mu _{1}}Q_{14}^{2}+e^{-2\mu _{2}}Q_{24}^{2}],
\end{eqnarray*}%
where the rest of coefficients are defined by similar formulas with a
corresponding changings of indices and partial derivative operators, $%
R_{22}, $ $R_{42}$ and $R_{32}$ is like $R_{11},R_{41}$ and $R_{31}$ with
with changing the index $1\rightarrow 2.$ The values $Q_{ij}$ and $Q_{i4}$
are defined respectively%
\begin{equation*}
Q_{ij}=\delta _{j}q_{i}-\delta _{i}q_{j}\mbox{ and
}Q_{i4}=\partial _{4}q_{i}-\delta _{i}\omega .
\end{equation*}

The nontrivial coefficients of the Einstein tensor are
\begin{eqnarray}
G_{11} &=&e^{-2\mu _{2}}[\delta _{22}^{2}(\mu _{3}+\mu _{4})+\delta _{2}(\mu
_{3}+\mu _{4})\delta _{2}(\mu _{4}-\mu _{2})+\delta _{2}\mu _{3}\delta
_{2}\mu _{3}]-  \notag \\
&&e^{-2\mu _{4}}[\partial _{44}^{2}(\mu _{3}+\mu _{2})+\partial _{4}(\mu
_{3}+\mu _{2})\partial _{4}(\mu _{2}-\mu _{4})+\partial _{4}\mu _{3}\partial
_{4}\mu _{3}]+  \notag \\
&&e^{-2\mu _{1}}[\delta _{1}\mu _{4}+\delta _{1}(\mu _{3}+\mu _{2})+\delta
_{1}\mu _{3}\delta _{1}\mu _{2}]-  \label{einstp} \\
&&\frac{1}{4}e^{2\mu _{3}}[e^{-2(\mu _{1}+\mu _{2})}Q_{12}^{2}-e^{-2(\mu
_{1}+\mu _{4})}Q_{14}^{2}+e^{-2(\mu _{2}+\mu _{3})}Q_{24}^{2}],  \notag
\end{eqnarray}%
\begin{eqnarray*}
G_{33} &=&e^{-2\mu _{1}}[\delta _{11}^{2}(\mu _{4}+\mu _{2})+\delta _{1}\mu
_{4}\delta _{1}(\mu _{4}-\mu _{1}+\mu _{2})+\delta _{1}\mu _{2}\delta
_{1}(\mu _{2}-\mu _{1})]+ \\
&&e^{-2\mu _{2}}[\delta _{22}^{2}(\mu _{4}+\mu _{1})+\delta _{2}(\mu
_{4}-\mu _{2}+\mu _{1})+\delta _{2}\mu _{1}\partial _{2}(\mu _{1}-\mu _{2})]-
\\
&&e^{-2\mu _{4}}[\partial _{44}^{2}(\mu _{1}+\mu _{2})+\partial _{4}\mu
_{1}\partial _{4}(\mu _{1}-\mu _{4})+\partial _{4}\mu _{2}\partial _{4}(\mu
_{2}-\mu _{4})+\partial _{4}\mu _{1}\partial _{4}\mu _{2}]+ \\
&&\frac{3}{4}e^{2\mu _{3}}[e^{-2(\mu _{1}+\mu _{2})}Q_{12}^{2}-e^{-2(\mu
_{1}+\mu _{4})}Q_{14}^{2}-e^{-2(\mu _{2}+\mu _{3})}Q_{24}^{2}],
\end{eqnarray*}%
\begin{eqnarray*}
G_{44} &=&e^{-2\mu _{1}}[\delta _{11}^{2}(\mu _{3}+\mu _{2})+\delta _{1}\mu
_{3}\delta _{1}(\mu _{3}-\mu _{1}+\mu _{2})+\delta _{1}\mu _{2}\delta
_{1}(\mu _{2}-\mu _{1})]- \\
&&e^{-2\mu _{2}}[\delta _{22}^{2}(\mu _{3}+\mu _{1})+\delta _{2}(\mu
_{3}-\mu _{2}+\mu _{1})+\delta _{2}\mu _{1}\partial _{2}(\mu _{1}-\mu _{2})]-%
\frac{1}{4}e^{2(\mu _{3}-\mu _{1}-\mu _{2})}Q_{12}^{2} \\
&&+e^{-2\mu _{4}}[\partial _{4}\mu _{3}\partial _{4}(\mu _{1}+\mu
_{2})+\partial _{4}\mu _{1}\partial _{4}\mu _{2}]-\frac{1}{4}e^{2(\mu
_{3}-\mu _{4})}[e^{-2\mu _{1}}Q_{14}^{2}-e^{-2\mu _{2}}Q_{24}^{2}].
\end{eqnarray*}%
The component $G_{22}$ is to be found from $G_{11}$ by changing the index $%
1\rightarrow 2.$ We note that the formulas (\ref{einstp}) transform into
similar ones from Ref. \cite{vels} if $\delta _{2}\rightarrow \partial _{2}.$

\subsection{Axial metric perturbations}

Axial perturbations are characterized by non--vanishing $\omega $ and $q_{i}$
which satisfy the equations
\begin{equation*}
R_{3i}=0,
\end{equation*}%
see the explicit formulas for such coefficients of the Ricci tensor in (\ref%
{riccip}). The resulting equations governing axial perturbations, $\delta
R_{31}=0,$ $\delta R_{32}=0,$ are respectively%
\begin{eqnarray}
\delta _{2}\left( e^{3\mu _{3}^{(\varepsilon )}+\mu _{4}^{(\varepsilon
)}-\mu _{1}^{(\varepsilon )}-\mu _{2}^{(\varepsilon )}}Q_{12}\right)
&=&-e^{3\mu _{3}^{(\varepsilon )}-\mu _{4}^{(\varepsilon )}-\mu
_{1}^{(\varepsilon )}+\mu _{2}^{(\varepsilon )}}\partial _{4}Q_{14},
\label{eq1} \\
\delta _{1}\left( e^{3\mu _{3}^{(\varepsilon )}+\mu _{4}^{(\varepsilon
)}-\mu _{1}^{(\varepsilon )}-\mu _{2}^{(\varepsilon )}}Q_{12}\right)
&=&e^{3\mu _{3}^{(\varepsilon )}-\mu _{4}^{(\varepsilon )}+\mu
_{1}^{(\varepsilon )}-\mu _{2}^{(\varepsilon )}}\partial _{4}Q_{24},  \notag
\end{eqnarray}%
where
\begin{equation}
Q_{ij}=\delta _{i}q_{j}-\delta _{j}q_{i},Q_{i4}=\partial _{4}q_{i}-\delta
_{i}\omega  \label{eq1a}
\end{equation}%
and for $\mu _{i}$ there are considered unperturbed values $\mu
_{i}^{(\varepsilon )}.$ Introducing the values of coefficients (\ref{coef1})
and (\ref{coef2}) \ and assuming that the perturbations have a time
dependence of type $\exp (i\sigma t)$ for a real constant $\sigma ,$ \ we
rewrite the equations (\ref{eq1})
\begin{eqnarray}
\frac{1+\varepsilon \left( \Delta ^{-1}+3r^{2}\phi /2\right) }{r^{4}\sin
^{3}\theta (\eta _{3}^{[v]})^{3/2}}\delta _{2}Q^{(\eta )} &=&-i\sigma \delta
_{r}\omega -\sigma ^{2}q_{1},  \label{eq2a} \\
\frac{\Delta }{r^{4}\sin ^{3}\theta (\eta _{3}^{[v]})^{3/2}}\delta
_{1}\left\{ Q^{(\eta )}\left[ 1+\frac{\varepsilon }{2}\left( \frac{\eta -1}{%
\Delta }-r^{2}\phi \right) \right] \right\} &=&i\sigma \partial _{\theta
}\omega +\sigma ^{2}q_{2}  \label{eq2b}
\end{eqnarray}%
for
\begin{equation*}
Q^{(\eta )}(r,\theta ,\varphi _{0},t)=\Delta Q_{12}\sin ^{3}\theta =\Delta
\sin ^{3}\theta (\partial _{2}q_{1}-\delta _{1}q_{2}),\Delta =r^{2}-2mr,
\end{equation*}%
where $\phi =0$ for solutions with $\Omega =1$ and $\phi (r,\varphi )=\eta
_{3}^{[v]}\left( r,\theta ,\varphi \right) \sin ^{2}\theta ,$ i. e. $\eta
_{3}\left( r,\theta ,\varphi \right) \sim \sin ^{-2}\theta $ for solutions
with $\Omega =1+\varepsilon ....$

We can exclude the function $\omega $ and define an equation for $Q^{(\eta
)} $ if we take the sum of the (\ref{eq2a}) subjected by the action of
operator $\partial _{2}$ and of the (\ref{eq2b}) subjected by the action of
operator $\delta _{1}.$ Using the relations (\ref{eq1a}), we write%
\begin{eqnarray*}
r^{4}\delta _{1}\left\{ \frac{\Delta }{r^{4}(\eta _{3}^{[v]})^{3/2}}\left[
\delta _{1}\left[ Q^{(\eta )}+\frac{\varepsilon }{2}\left( \frac{\eta -1}{%
\Delta }-r^{2}\right) \phi \right] \right] \right\} + && \\
\sin ^{3}\theta \partial _{2}\left[ \frac{1+\varepsilon (\Delta
^{-1}+3r^{2}\phi /2)}{\sin ^{3}\theta (\eta _{3}^{[v]})^{3/2}}\delta
_{2}Q^{(\eta )}\right] +\frac{\sigma ^{2}r^{4}}{\Delta \eta _{3}^{3/2}}%
Q^{(\eta )} &=&0.
\end{eqnarray*}%
The solution of this equation is searched in the form $Q^{(\eta
)}=Q+\varepsilon Q^{(1)}$ which results in
\begin{equation}
r^{4}\partial _{1}\left( \frac{\Delta }{r^{4}(\eta _{3}^{[v]})^{3/2}}%
\partial _{1}Q\right) +\sin ^{3}\theta \partial _{2}\left( \frac{1}{\sin
^{3}\theta (\eta _{3}^{[v]})^{3/2}}\partial _{2}Q\right) +\frac{\sigma
^{2}r^{4}}{\Delta (\eta _{3}^{[v]})^{3/2}}Q=\varepsilon A\left( r,\theta
,\varphi _{0}\right) ,  \label{eq3}
\end{equation}%
where%
\begin{eqnarray*}
A\left( r,\theta ,\varphi _{0}\right) &=&r^{4}\partial _{1}\left( \frac{%
\Delta }{r^{4}(\eta _{3}^{[v]})^{3/2}}n_{1}\right) \frac{\partial Q}{%
\partial t}-r^{4}\partial _{1}\left( \frac{\Delta }{r^{4}(\eta
_{3}^{[v]})^{3/2}}\partial _{1}Q^{(1)}\right) \\
&&-\sin ^{3}\theta \delta _{2}\left[ \frac{1+\varepsilon (\Delta
^{-1}+3r^{2}\phi /2)}{\sin ^{3}\theta (\eta _{3}^{[v]})^{3/2}}\delta
_{2}Q^{(1)}-\frac{\sigma ^{2}r^{4}}{\Delta (\eta _{3}^{[v]})^{3/2}}Q^{(1)}%
\right] ,
\end{eqnarray*}%
with a time dependence like $\exp [i\sigma t]$

It is possible to construct different classes of solutions of the equation (%
\ref{eq3}). At the first step we find the solution for $Q$ when $\varepsilon
=0.$ Then, for a known value of $Q\left( r,\theta ,\varphi _{0}\right) $
from
\begin{equation*}
Q^{(\eta )}=Q+\varepsilon Q^{(1)},
\end{equation*}%
we can define $Q^{(1)}$ from the equations (\ref{eq2a}) and (\ref{eq2b}) by
considering the values proportional to $\varepsilon $ which can be written
\begin{eqnarray}
\partial _{1}Q^{(1)} &=&B_{1}\left( r,\theta ,\varphi _{0}\right) ,
\label{eq4} \\
\partial _{2}Q^{(1)} &=&B_{2}\left( r,\theta ,\varphi _{0}\right) .  \notag
\end{eqnarray}%
The integrability condition of the system (\ref{eq4}), $\partial
_{1}B_{2}=\partial _{2}B_{1}$ imposes a relation between the polarization
functions $\eta _{3},\eta ,w_{1}$and $n_{1}$ (for a corresponding class of
solutions). In order to prove that there are stable anisotropic
configurations of anisotropic black hole solutions, we may consider a set of
polarization functions when $A\left( r,\theta ,\varphi _{0}\right) =0$ and
the solution with $Q^{(1)}=0$ is admitted. This holds, for example, if
\begin{equation*}
\Delta n_{1}=n_{0}r^{4}(\eta _{3}^{[v]})^{3/2},\ n_{0}=const.
\end{equation*}%
In this case the axial perturbations are described by the equation
\begin{equation}
(\eta _{3}^{[v]})^{3/2}r^{4}\partial _{1}\left( \frac{\Delta }{r^{4}(\eta
_{3}^{[v]})^{3/2}}\partial _{1}Q\right) +\sin ^{3}\theta \delta _{2}\left(
\frac{1}{\sin ^{3}\theta }\delta _{2}Q\right) +\frac{\sigma ^{2}r^{4}}{%
\Delta }Q=0  \label{eq5}
\end{equation}%
which is obtained from (\ref{eq3}) for $\eta _{3}^{[v]}=\eta
_{3}^{[v]}\left( r,\varphi _{0}\right) ,$ or for $\phi (r,\varphi _{0})=\eta
_{3}^{[v]}\left( r,\theta ,\varphi _{0}\right) \sin ^{2}\theta .$

In the limit $\eta _{3}^{[v]}\rightarrow 1$ the solution of equation (\ref%
{eq5}) is investigated in details in Ref. \cite{chan}. \ Here, we prove that
in a similar manner we can define exact solutions for non--trivial values of
$\eta _{3}^{[v]}.$ \ The variables $r$ and $\theta $ can be separated if we
substitute
\begin{equation*}
Q(r,\theta ,\varphi _{0})=Q_{0}(r,\varphi _{0})C_{l+2}^{-3/2}(\theta ),
\end{equation*}%
where $C_{n}^{\nu }$ are the Gegenbauer functions generated by the equation%
\begin{equation*}
\left[ \frac{d}{d\theta }\sin ^{2\nu }\theta \frac{d}{d\theta }+n\left(
n+2\nu \right) \sin ^{2\nu }\theta \right] C_{n}^{\nu }(\theta )=0.
\end{equation*}%
The function $C_{l+2}^{-3/2}(\theta )$ is related to the second derivative
of the Legendre function $P_{l}(\theta )$ by formulas%
\begin{equation*}
C_{l+2}^{-3/2}(\theta )=\sin ^{3}\theta \frac{d}{d\theta }\left[ \frac{1}{%
\sin \theta }\frac{dP_{l}(\theta )}{d\theta }\right] .
\end{equation*}%
The separated part of (\ref{eq5}) depending on radial variable with a fixed
value $\varphi _{0}$ transforms into the equation%
\begin{equation}
(\eta _{3}^{[v]})^{3/2}\Delta \frac{d}{dr}\left( \frac{\Delta }{r^{4}(\eta
_{3}^{[v]})^{3/2}}\frac{dQ_{0}}{dr}\right) +\left( \sigma ^{2}-\frac{\mu
^{2}\Delta }{r^{4}}\right) Q_{0}=0,  \label{eq6}
\end{equation}%
where $\mu ^{2}=(l-1)(l+2)$ for $l=2,3,...$ A further simplification is
possible for $\eta _{3}^{[v]}=\eta _{3}^{[v]}(r,\varphi _{0})$ if we
introduce in the equation (\ref{eq6}) a new radial coordinate
\begin{equation*}
r_{\#}=\int (\eta _{3}^{[v]})^{3/2}(r,\varphi _{0})r^{2}dr
\end{equation*}%
and a new unknown function $Z^{(\eta )}=r^{-1}Q_{0}(r).$ The equation for $%
Z^{(\eta )}$ is an Schrodinger like one--dimensional wave equation%
\begin{equation}
\left( \frac{d^{2}}{dr_{\#}^{2}}+\frac{\sigma ^{2}}{(\eta _{3}^{[v]})^{3/2}}%
\right) Z^{(\eta )}=V^{(\eta )}Z^{(\eta )}  \label{eq7}
\end{equation}%
with the potential
\begin{equation}
V^{(\eta )}=\frac{\Delta }{r^{5}(\eta _{3}^{[v]})^{3/2}}\left[ \mu ^{2}-r^{4}%
\frac{d}{dr}\left( \frac{\Delta }{r^{4}(\eta _{3}^{[v]})^{3/2}}\right) %
\right]  \label{eq7a}
\end{equation}%
and polarized parameter
\begin{equation*}
\widetilde{\sigma }^{2}=\sigma ^{2}/(\eta _{3}^{[v]})^{3/2}.
\end{equation*}%
This equation transforms into the so--called Regge--Wheeler equation if $%
\eta _{3}^{[v]}=1.$ For instance, for the Schwarzschild black hole such
solutions are investigated and tabulated for different values of $l=2,3$ and
$4$ in Ref. \cite{chan}.

We note that for static anisotropic black holes with nontrivial anisotropic
conformal factor, $\Omega =1+\varepsilon ...,$ even $\eta _{3}$ may depend
on angular variable $\theta $ because of condition that $\phi (r,\varphi
_{0})=\eta _{3}^{[v]}\left( r,\theta ,\varphi _{0}\right) \sin ^{2}\theta $
the equation (\ref{eq5}) transforms directly in (\ref{eq7}) with $\mu =0$
without any separation of variables $r$ and $\theta .$ It is not necessary
in this case to consider the Gegenbauer functions because $Q_{0}$ does not
depend on $\theta $ which corresponds to a solution with $l=1.$

We may transform (\ref{eq7}) into the usual form,
\begin{equation*}
\left( \frac{d^{2}}{dr_{\star }^{2}}+\sigma ^{2}\right) Z^{(\eta )}=%
\widetilde{V}^{(\eta )}Z^{(\eta )}
\end{equation*}%
if we introduce the variable
\begin{equation*}
r_{\star }=\int dr_{\#}(\eta _{3}^{[v]})^{-3/2}\left( r_{\#},\varphi
_{0}\right)
\end{equation*}%
for $\widetilde{V}^{(\eta )}=(\eta _{3}^{[v]})^{3/2}V^{(\eta )}.$ So, the
polarization function $\eta _{3}^{[v]},$ describing static anholonomic
deformations of the Scharzshild black hole, ''renormalizes'' the potential
in the one--dimensional Schrodinger wave--equation governing axial
perturbations of such objects.

We conclude that small static ''ellipsoid'' like deformations and
polarizations of constants of spherical black holes (the anisotropic
configurations being described by generic off--diagonal metric ansatz) do
not change the type of equations for axial petrubations: one modifies the
potential barrier,%
\begin{equation*}
V^{(-)}=\frac{\Delta }{r^{5}}\left[ \left( \mu ^{2}+2\right) r-6m\right]
\longrightarrow \widetilde{V}^{(\eta )}
\end{equation*}%
and re--defines the radial variables%
\begin{equation*}
r_{\ast }=r+2m\ln \left( r/2m-1\right) \longrightarrow r_{\star }(\varphi
_{0})
\end{equation*}%
with a parametric dependence on anisotropic angular coordinate which is
caused by the existence of a deformed static horizon.

\subsection{Polar metric perturbations}

The polar perturbations are described by non--trivial increments of the
diagonal metric coefficients, $\delta \mu _{\alpha }=\delta \mu _{\alpha
}^{(\varepsilon )}+\delta \mu _{\alpha }^{(\varsigma )},$ for
\begin{equation*}
\mu _{\alpha }^{(\varepsilon )}=\nu _{\alpha }+\delta \mu _{\alpha
}^{(\varepsilon )}
\end{equation*}%
where $\delta \mu _{\alpha }^{(\varsigma )}(x^{k},t)$ parametrize time
depending fluctuations which are stated to be the same both for spherical
and/or spheroid configurations and $\delta \mu _{\alpha }^{(\varepsilon )}$
is a static deformation from the spherical symmetry. Following notations (%
\ref{coef1}) and (\ref{coef2}) we write%
\begin{equation*}
e^{v_{1}}=r/\sqrt{|\Delta |},e^{v_{2}}=r\sqrt{|q^{[v]}(r)|}%
,e^{v_{3}}=rh_{3}\sin \theta ,e^{v_{4}}=\Delta /r^{2}
\end{equation*}%
and
\begin{equation*}
\delta \mu _{1}^{(\varepsilon )}=-\frac{\varepsilon }{2}\left( \Delta
^{-1}+r^{2}\phi \right) ,\delta \mu _{2}^{(\varepsilon )}=\delta \mu
_{3}^{(\varepsilon )}=-\frac{\varepsilon }{2}r^{2}\phi ,\delta \mu
_{4}^{(\varepsilon )}=\frac{\varepsilon \eta }{2\Delta }
\end{equation*}%
where $\phi =0$ for the solutions with $\Omega =1.$

Examining the expressions for $R_{4i},R_{12,}R_{33}$ and $G_{11}$ (see
respectively (\ref{riccip}) and (\ref{einstp})\ ) we conclude that the
values $Q_{ij}$ appear quadratically which can be ignored in a linear
perturbation theory. Thus the equations for the axial and the polar
perturbations decouple. Considering only linearized expressions, both for
static $\varepsilon $--terms and fluctuations depending on time about the
Schwarzschild values we obtain the equations%
\begin{eqnarray}
\delta _{1}\left( \delta \mu _{2}+\delta \mu _{3}\right) +\left(
r^{-1}-\delta _{1}\mu _{4}\right) \left( \delta \mu _{2}+\delta \mu
_{3}\right) -2r^{-1}\delta \mu _{1} &=&0\quad \left( \delta R_{41}=0\right) ,
\notag \\
\delta _{2}\left( \delta \mu _{1}+\delta \mu _{3}\right) +\left( \delta \mu
_{2}-\delta \mu _{3}\right) \cot \theta &=&0\quad \left( \delta
R_{42}=0\right) ,  \notag \\
\delta _{2}\delta _{1}\left( \delta \mu _{3}+\delta \mu _{4}\right) -\delta
_{1}\left( \delta \mu _{2}-\delta \mu _{3}\right) \cot \theta - &&  \notag \\
\left( r^{-1}-\delta _{1}\mu _{4}\right) \delta _{2}(\delta \mu _{4})-\left(
r^{-1}+\delta _{1}\mu _{4}\right) \delta _{2}(\delta \mu _{1}) &=&0\quad
\left( \delta R_{42}=0\right) ,  \notag \\
e^{2\mu _{4}}\{2\left( r^{-1}+\delta _{1}\mu _{4}\right) \delta _{1}(\delta
\mu _{3})+r^{-1}\delta _{1}\left( \delta \mu _{3}+\delta \mu _{4}-\delta \mu
_{1}+\delta \mu _{2}\right) + &&  \label{peq1} \\
\delta _{1}\left[ \delta _{1}(\delta \mu _{3})\right] -2r^{-1}\delta \mu
_{1}\left( r^{-1}+2\delta _{1}\mu _{4}\right) \}-2e^{-2\mu _{4}}\partial
_{4}[\partial _{4}(\delta \mu _{3})]+ &&  \notag \\
r^{-2}\{\delta _{2}[\delta _{2}(\delta \mu _{3})]+\delta _{2}\left( 2\delta
\mu _{3}+\delta \mu _{4}+\delta \mu _{1}-\delta \mu _{2}\right) \cot \theta
+2\delta \mu _{2}\} &=&0\quad \left( \delta R_{33}=0\right) ,  \notag \\
e^{-2\mu _{1}}[r^{-1}\delta _{1}(\delta \mu _{4})+\left( r^{-1}+\delta
_{1}\mu _{4}\right) \delta _{1}\left( \delta \mu _{2}+\delta \mu _{3}\right)
- &&  \notag \\
2r^{-1}\delta \mu _{1}\left( r^{-1}+2\delta _{1}\mu _{4}\right) ]-e^{-2\mu
_{4}}\partial _{4}[\partial _{4}(\delta \mu _{3}+\delta \mu _{2})] &&  \notag
\\
+r^{-2}\{\delta _{2}[\delta _{2}(\delta \mu _{3})]+\delta _{2}\left( 2\delta
\mu _{3}+\delta \mu _{4}-\delta \mu _{2}\right) \cot \theta +2\delta \mu
_{2}\} &=&0\quad \left( \delta G_{11}=0\right) .  \notag
\end{eqnarray}

The values of type $\delta \mu _{\alpha }=\delta \mu _{\alpha
}^{(\varepsilon )}+\delta \mu _{\alpha }^{(\varsigma )}$ from (\ref{peq1})
contain two components: the first ones are static, proportional to $%
\varepsilon ,$ and the second ones may depend on time coordinate $t.$ We
shall assume that the perturbations $\delta \mu _{\alpha }^{(\varsigma )}$
have a time--dependence $\exp [\sigma t]$ $\ $so that the partial time
derivative $"\partial _{4}"$ is replaced by the factor $i\sigma .$ In order
to treat both type of increments in a similar fashion we may consider that
the values labeled with $(\varepsilon )$ also oscillate in time like $\exp
[\sigma ^{(\varepsilon )}t]$ but with a very small (almost zero) frequency $%
\sigma ^{(\varepsilon )}\rightarrow 0.$ There are also actions of
''elongated'' partial derivative operators like
\begin{equation*}
\delta _{1}\left( \delta \mu _{\alpha }\right) =\partial _{1}\left( \delta
\mu _{\alpha }\right) -\varepsilon n_{1}\partial _{4}\left( \delta \mu
_{\alpha }\right) .
\end{equation*}%
To avoid a calculus with complex values we associate the terms proportional $%
\varepsilon n_{1}\partial _{4}$ to amplitudes of type $\varepsilon
in_{1}\partial _{4}$ and write this operator as
\begin{equation*}
\delta _{1}\left( \delta \mu _{\alpha }\right) =\partial _{1}\left( \delta
\mu _{\alpha }\right) +\varepsilon n_{1}\sigma \left( \delta \mu _{\alpha
}\right) .
\end{equation*}%
For the ''non-perturbed'' Schwarzschild values, which are static, the
operator $\delta _{1}$ reduces to $\partial _{1},$ i.e. $\delta
_{1}v_{\alpha }=\partial _{1}v_{\alpha }.$ Hereafter we shall consider that
the solution of the system (\ref{peq1}) consists \ from a superposition of
two linear solutions, $\delta \mu _{\alpha }=\delta \mu _{\alpha
}^{(\varepsilon )}+\delta \mu _{\alpha }^{(\varsigma )}$; the first class of
solutions for increments will be provided with index $(\varepsilon ),$
corresponding to the frequence $\sigma ^{(\varepsilon )}$ and the second
class will be for the increments with index $(\varsigma )$ and correspond to
the frequence $\sigma ^{(\varsigma )}.$ We shall write this as $\delta \mu
_{\alpha }^{(A)}$ and $\sigma _{(A)}$ for the labels $A=\varepsilon $ or $%
\varsigma $ and suppress the factors $\exp [\sigma ^{(A)}t]$ in our
subsequent considerations. The system of equations (\ref{peq1}) will be
considered for both type of increments.

We can separate the variables by substitutions (see the method in Refs. \cite%
{fried,chan})%
\begin{eqnarray}
\delta \mu _{1}^{(A)} &=&L^{(A)}(r)P_{l}(\cos \theta ),\quad \delta \mu
_{2}^{(A)}=\left[ T^{(A)}(r)P_{l}(\cos \theta )+V^{(A)}(r)\partial
^{2}P_{l}/\partial \theta ^{2}\right] ,  \label{auxc1} \\
\delta \mu _{3}^{(A)} &=&\left[ T^{(A)}(r)P_{l}(\cos \theta )+V^{(A)}(r)\cot
\theta \partial P_{l}/\partial \theta \right] ,\quad \delta \mu
_{4}^{(A)}=N^{(A)}(r)P_{l}(\cos \theta )  \notag
\end{eqnarray}%
and reduce the system of equations (\ref{peq1}) \ to%
\begin{eqnarray}
\delta _{1}\left( N^{(A)}-L^{(A)}\right) &=&\left( r^{-1}-\partial _{1}\nu
_{4}\right) N^{(A)}+\left( r^{-1}+\partial _{1}\nu _{4}\right) L^{(A)},
\notag \\
\delta _{1}L^{(A)}+\left( 2r^{-1}-\partial _{1}\nu _{4}\right) N^{(A)} &=&-
\left[ \delta _{1}X^{(A)}+\left( r^{-1}-\partial _{1}\nu _{4}\right) X^{(A)}%
\right] ,  \label{peq2a}
\end{eqnarray}%
and
\begin{eqnarray}
2r^{-1}\delta _{1}\left( N^{(A)}\right)
-l(l+1)r^{-2}e^{-2v_{4}}N^{(A)}-2r^{-1}(r^{-1}+2\partial _{1}\nu
_{4})L^{(A)}-2(r^{-1}+ &&  \label{peq2b} \\
\partial _{1}\nu _{4})\delta _{1}\left[ N^{(A)}+(l-1)(l+2)V^{(A)}/2\right]
-(l-1)(l+2)r^{-2}e^{-2v_{4}}\left( V^{(A)}-L^{(A)}\right) - &&  \notag \\
2\sigma _{(A)}^{2}e^{-4v_{4}}\left[ L^{(A)}+(l-1)(l+2)V^{(A)}/2\right] &=&0,
\notag
\end{eqnarray}%
where we have introduced new functions
\begin{equation*}
X^{(A)}=\frac{1}{2}(l-1)(l+2)V^{(A)}
\end{equation*}%
and considered the relation
\begin{equation*}
T^{(A)}-V^{(A)}+L^{(A)}=0\quad (\delta R_{42}=0).
\end{equation*}%
We can introduce the functions
\begin{eqnarray}
\widetilde{L}^{(A)} &=&L^{(A)}+\varepsilon \sigma _{(A)}\int
n_{1}L^{(A)}dr,\quad \widetilde{N}^{(A)}=N^{(A)}+\varepsilon \sigma
_{(A)}\int n_{1}N^{(A)}dr,  \label{tilds} \\
\widetilde{T}^{(A)} &=&N^{(A)}+\varepsilon \sigma _{(A)}\int
n_{1}N^{(A)}dr,\quad \widetilde{V}^{(A)}=V^{(A)}+\varepsilon \sigma
_{(A)}\int n_{1}V^{(A)}dr,  \notag
\end{eqnarray}%
for which
\begin{equation*}
\partial _{1}\widetilde{L}^{(A)}=\delta _{1}\left( L^{(A)}\right) ,\partial
_{1}\widetilde{N}^{(A)}=\delta _{1}\left( N^{(A)}\right) ,\partial _{1}%
\widetilde{T}^{(A)}=\delta _{1}\left( T^{(A)}\right) ,\partial _{1}%
\widetilde{V}^{(A)}=\delta _{1}\left( V^{(A)}\right) ,
\end{equation*}%
and, this way it is possible to substitute in (\ref{peq2a}) and (\ref{peq2b}%
) the elongated partial derivative $\delta _{1}$ by the usual one acting on
''tilded'' radial increments.

By straightforward calculations (see details in Ref. \cite{chan}) one can
check that the functions
\begin{equation*}
Z_{(A)}^{(+)}=r^{2}\frac{6mX^{(A)}/r(l-1)(l+2)-L^{(A)}}{r(l-1)(l+2)/2+3m}
\end{equation*}%
satisfy one--dimensional wave equations similar to (\ref{eq7}) for $Z^{(\eta
)}$ with $\eta _{3}=1,$ when $r_{\star }=r_{\ast },$%
\begin{eqnarray}
\left( \frac{d^{2}}{dr_{\ast }^{2}}+\sigma _{(A)}^{2}\right) \widetilde{Z}%
_{(A)}^{(+)} &=&V^{(+)}Z_{(A)}^{(+)},  \label{peq3} \\
\widetilde{Z}_{(A)}^{(+)} &=&Z_{(A)}^{(+)}+\varepsilon \sigma _{(A)}\int
n_{1}Z_{(A)}^{(+)}dr,  \notag
\end{eqnarray}%
where%
\begin{eqnarray}
V^{(+)} &=&\frac{2\Delta }{r^{5}[r(l-1)(l+2)/2+3m]^{2}}\times \{9m^{2}\left[
\frac{r}{2}(l-1)(l+2)+m\right]  \label{pot3} \\
&&+\frac{1}{4}(l-1)^{2}(l+2)^{2}r^{3}\left[ 1+\frac{1}{2}(l-1)(l+2)+\frac{3m%
}{r}\right] \}.  \notag
\end{eqnarray}%
For $\varepsilon \rightarrow 0,$ the equation (\ref{peq3}) transforms in the
usual Zerilli equation \cite{zerilli,chan}.

To complete the solution we give the formulas for the ''tilded'' $L$--, $X$%
-- and $N$--factors,%
\begin{eqnarray}
\widetilde{L}^{(A)} &=&\frac{3m}{r^{2}}\widetilde{\Phi }^{(A)}-\frac{%
(l-1)(l+2)}{2r}\widetilde{Z}_{(A)}^{(+)},  \label{peg3a} \\
\widetilde{X}^{(A)} &=&\frac{(l-1)(l+2)}{2r}(\widetilde{\Phi }^{(A)}+%
\widetilde{Z}_{(A)}^{(+)}),  \notag \\
\widetilde{N}^{(A)} &=&\left( m-\frac{m^{2}+r^{4}\sigma _{(A)}^{2}}{r-2m}%
\right) \frac{\widetilde{\Phi }^{(A)}}{r^{2}}-\frac{(l-1)(l+2)r}{%
2(l-1)(l+2)+12m}\frac{\partial \widetilde{Z}_{(A)}^{(+)}}{\partial r_{\#}}
\notag \\
&&-\frac{(l-1)(l+2)}{\left[ r(l-1)(l+2)+6m\right] ^{2}}\times  \notag \\
&&\left\{ \frac{12m^{2}}{r}+3m(l-1)(l+2)+\frac{r}{2}(l-1)(l+2)\left[
(l-1)(l+2)+2\right] \right\} ,  \notag
\end{eqnarray}%
where
\begin{equation*}
\widetilde{\Phi }^{(A)}=(l-1)(l+2)e^{\nu _{4}}\int \frac{e^{-\nu _{4}}%
\widetilde{Z}_{(A)}^{(+)}}{(l-1)(l+2)r+6m}dr.
\end{equation*}%
Following \ the relations (\ref{tilds}) we can compute the corresponding
''untileded'' values an put them in (\ref{auxc1}) in oder to find the
increments of fluctuations driven by the system of equations (\ref{peq1}).
For simplicity, we omit the rather compersome final expressions.

The formulas (\ref{peg3a}) together with a solution of the wave equation (%
\ref{peq3}) complete the procedure of definition of formal solutions for
polar perturbations. In Ref. \cite{chan} there are tabulated the data for
the potential (\ref{pot3}) for different values of $l$ and $\left(
l-1\right) (l+2)/2.$ In the anisotropic case the explicit form of solutions
is deformed by terms proportional to $\varepsilon n_{1}\sigma .$ The static
ellipsoidal like deformations can be modeled by the formulas obtained in the
limit $\sigma _{(\varepsilon )}\rightarrow 0.$

\subsection{The stability of polarized black ellipsoids}

The problem of stability of anholonomically deformed Schwarzschild metrics
to external perturbation is very important to be solved in order to
understand if such static black ellipsoid like objects may exist in general
relativity and its cosmological constant generalizations. We address the
question: Let be given any initial values for a static locally anisotropic
configuration confined to a finite interval of $r_{\star },$ for axial
perturbations, and $r_{\ast },$ for polar peturbations, will one remain
bounded such peturbations at all times of evolution? The answer to this
question is to obtained symilarly to Refs. \cite{chan} and \cite{vels} with
different type of definitions of functions g $Z^{(\eta )}$ and $%
Z_{(A)}^{(+)} $ for different type of black holes.

We have proved that even for anisotropic configurations every type of
perturbations are governed by one dimensional wave equations of the form%
\begin{equation}
\frac{d^{2}Z}{d\rho }+\sigma ^{2}Z=VZ  \label{eq8}
\end{equation}%
where $\rho $ is a radial type coordinate, $Z$ is a corresponding $Z^{(\eta
)}$ or $Z_{(A)}^{(+)}$ with respective smooth real, independent of $\sigma
>0 $ potentials $\widetilde{V}^{(\eta )}$ or $V^{(-)}$ with bounded
integrals. For such equations a solution $Z(\rho ,\sigma ,\varphi _{0})$
satisfying the boundary conditions \ $Z\rightarrow e^{i\sigma \rho
}+R(\sigma )e^{-i\sigma \rho }\quad (\rho \rightarrow +\infty )$ and $%
Z\rightarrow T(\sigma )e^{i\sigma \rho }\qquad (\rho \rightarrow -\infty )$
(the first expression corresponds to an incident wave of unit amplitude from
$+\infty $ giving rise to a reflected wave of amplitude $R(\sigma )$ at $%
+\infty $ and the second expression is for a transmitted wave of \ amplitude
$T(\sigma )$ at $-\infty ),$ provides a basic complete set of wave functions
which allows to obtain a stable evolution. For any initial perturbation that
is smooth and confined to finite interval of $\rho ,$ we can write the
integral%
\begin{equation*}
\psi (\rho ,0)=\left( 2\pi \right) ^{-1/2}\int_{-\infty }^{+\infty }\widehat{%
\psi }(\sigma ,0)Z(\rho ,\sigma )d\sigma
\end{equation*}%
and define the evoluiton of perturbations,%
\begin{equation*}
\psi (\rho ,t)=\left( 2\pi \right) ^{-1/2}\int_{-\infty }^{+\infty }\widehat{%
\psi }(\sigma ,0)e^{i\sigma t}Z(\rho ,\sigma )d\sigma .
\end{equation*}%
The Schrodinger theory garantees the conditions%
\begin{equation*}
\int_{-\infty }^{+\infty }|\psi (\rho ,0)|^{2}d\rho =\int_{-\infty
}^{+\infty }|\widehat{\psi }(\sigma ,0)|^{2}d\sigma =\int_{-\infty
}^{+\infty }|\psi (\rho ,0)|^{2}d\rho ,
\end{equation*}%
from which the boundedness of $\psi (\rho ,t)$ follows for all $t>0.$

In our consideration we have replaced the time partial derivative $\partial
/\partial t$ by $i\sigma ,$ which was represented by the approximation of
perturbations to be periodic like $e^{i\sigma t}.$ This is connected with a
time--depending variant of (\ref{eq8}), like%
\begin{equation*}
\frac{\partial ^{2}Z}{\partial t^{2}}=\frac{\partial ^{2}Z}{\partial \rho
^{2}}-VZ.
\end{equation*}%
Multiplying this equation on $\partial \overline{Z}/\partial t,$ where $%
\overline{Z}$ denotes the complex conjugation, and integrating on parts, we
obtain
\begin{equation*}
\int_{-\infty }^{+\infty }\left( \frac{\partial \overline{Z}}{\partial t}%
\frac{\partial ^{2}Z}{\partial t^{2}}+\frac{\partial Z}{\partial \rho }\frac{%
\partial ^{2}\overline{Z}}{\partial t\partial \rho }+VZ\frac{\partial
\overline{Z}}{\partial t}\right) d\rho =0
\end{equation*}%
providing the conditions of convergence of necessary integrals. This
equation added to its complex conjugate results in a constant energy
integral,
\begin{equation*}
\int_{-\infty }^{+\infty }\left( \left| \frac{\partial Z}{\partial t}\right|
^{2}+\left| \frac{\partial Z}{\partial \rho }\right| ^{2}+V\left| Z\right|
^{2}\right) d\rho =const,
\end{equation*}%
which bounds the expression $|\partial Z/\partial t|^{2}$ and excludes an
exponential growth of any bounded solution of the equation (\ref{eq8}). We
note that this property holds for every type of ''ellipsoidal'' like
deformation of the potential, $V\rightarrow V+\varepsilon V^{(1)},$ with
possible dependencies on polarization functions as we considered in (\ref%
{eq7a}) and/or (\ref{pot3}).

The general properties of the one--dimensional Schrodinger equations related
to perturbations of holonomic and anholonomic solutions of the Einstein
equations allow us to conclude that there are locally anisotropic static
configuratios which are stable under linear deformations.

In a similar manner we may analyze perturbations (axial or polar) governed
by a two--dimensional Schrodinger waive equation like
\begin{equation*}
\frac{\partial ^{2}Z}{\partial t^{2}}=\frac{\partial ^{2}Z}{\partial \rho
^{2}}+A(\rho ,\varphi ,t)\frac{\partial ^{2}Z}{\partial \varphi ^{2}}-V(\rho
,\varphi ,t)Z
\end{equation*}%
for some functions of necessary smooth class. The stability in this case is
proven if exists an (energy) integral
\begin{equation*}
\int_{0}^{\pi }\int_{-\infty }^{+\infty }\left( \left| \frac{\partial Z}{%
\partial t}\right| ^{2}+\left| \frac{\partial Z}{\partial \rho }\right|
^{2}+\left| A\frac{\partial Z}{\partial \rho }\right| ^{2}+V\left| Z\right|
^{2}\right) d\rho d\varphi =const
\end{equation*}%
which bounds $|\partial Z/\partial t|^{2}$ $\ $for two--dimensional
perturbations. For simplicity, we omitted such calculus in this work.

We emphasize that this way we can also prove the stability of perturbations
along ''aniso\-tro\-pic'' directions of arbitrary anholonomic deformations
of the Schwarzschild solution which have non--spherical horizons and can be
covered by a set of finite regions approximated as small, ellipsoid like,
deformations of some spherical hypersurfaces. We may analyze the geodesic
congruence on every deformed sub-region of necessary smoothly class and
proof the stability as we have done for the resolution ellipsoid horizons.
In general, we may consider horizons of with non--trivial topology, like
vacuum black tori, or higher genus anisotropic configurations. This is not
prohibited by the principles of topological censorship \cite{ptc} if we are
dealing with off--diagonal metrics and associated anholonomic frames \cite{v}%
. The vacuum anholonomy in such cases may be treated as an effective matter
which change the conditions of topological theorems.

\section{Two Additional Examples of Off--Diagonal Exact Solutions}

There are some classes of exact solutions wich can be modeled by anholonomic
frame transforms and generic off--diagonal metric ansatz and related to
configurations constructed by using another methods \cite{es,cans}. We
analyze in this section two classes of such 4D spacetimes.

\subsection{Anholonomic ellipsoidal shapes}

The present status of ellipsoidal shapes in general relativity associated to
some perfect--fluid bodies, rotating configurations or to some families of
confocal ellipsoids in Reimannian spaces is examined in details in Ref. \cite%
{es}. We shall illustrate in this subsection how such configurations may be
modeled by generic off--diagonal metrics and/or as spacetimes with
anisotropic cosmological constant. The off--diagonal coefficients will be
subjected to certain anholonomy conditions resulting (roughly speaking) in
effects similar to those of perfect--fluid bodies.

We consider a metric ansatz with conformal factor like in (\ref{metric1p})%
\begin{eqnarray}
\delta s^{2} &=&\Omega (\theta ,\nu )\left[ g_{1}d\theta ^{2}+g_{2}(\theta
)d\varphi ^{2}+h_{3}(\theta ,\nu )\delta \nu ^{2}+h_{4}(\theta ,\nu )\delta
t^{2}\right] ,  \notag \\
\delta \nu &=&d\nu +w_{1}(\theta ,\nu )d\theta +w_{2}(\theta ,\nu )d\varphi ,
\label{ansatzes} \\
\delta t &=&dt+n_{1}(\theta ,\nu )d\theta +n_{2}(\theta ,\nu )d\varphi ,
\notag
\end{eqnarray}%
where the coordinated $\left( x^{1}=\theta ,x^{2}=\varphi \right) $ are
holonomic and the coordinate $y^{3}=\nu $ and the timelike coordinate $%
y^{4}=t$ are 'anisotropic' ones. For a particular parametrization when
\begin{eqnarray}
\Omega &=&\Omega _{\lbrack 0]}(\nu )=v\left( \rho \right) \rho ^{2},\
g_{1}=1,  \label{dataes} \\
g_{2} &=&g_{2[0]}=\sin ^{2}\theta ,h_{3}=h_{3[0]}=1,h_{4}=h_{4[0]}=-1,
\notag \\
w_{1} &=&0,\ w_{2}=w_{2[0]}(\theta )=\sin ^{2}\theta ,\ n_{1}=0,\
n_{2}=n_{2[0]}(\theta )=2R_{0}\cos \theta  \notag
\end{eqnarray}%
and the coordinate $\nu $ is defined related to $\rho $ as
\begin{equation*}
d\nu =\int \left| \frac{f\left( \rho \right) }{v\left( \rho \right) }\right|
^{1/2}\frac{d\rho }{\rho },
\end{equation*}%
we obtain the metric element for a special case spacetimes with co--moving
ellipsoidal symmetry defined by an axially symmetric, rigidly rotating
perfect--fluid configuration with confocal inside ellipsoidal symmetry (see
formula (4.21) and related discussion in Ref. \cite{es}, where the status of
constant $R_{0}$ and functions $v\left( \rho \right) $ and $f\left( \rho
\right) $ are explicitly defined).

By introducing nontrivial ''polarization'' functions $q^{[v]}\left( \theta
\right) $ and $\eta _{3,4}(\theta ,\nu )$ for which
\begin{equation*}
g_{2}=g_{2[0]}q^{[v]}\left( \theta \right) ,\ h_{3,4}=\eta _{3,4}(\theta
,\nu )h_{3,4[0]}
\end{equation*}%
we can state the conditions when the ansatz (\ref{ansatzes}) defines a) an
off--diagonal ellipsoidal shape or b) an ellipsoidal configuration induced
by anisotropically polarized cosmological constant.

Let us consider the case a). The Theorem 2 from Ref. \cite{vth} and the
formula (72) in Ref. \cite{vmag2} (see also  the Appendix in \cite{vncs})
states that any metric of type (\ref{ansatzes}) is vacuum if $\Omega
^{p_{1}/p_{2}}=h_{3}$ for some integers $p_{1}$ and $p_{2},$ the factor $%
\Omega =\Omega _{\lbrack 1]}(\theta ,\nu )\Omega _{\lbrack 0]}(\nu )$
satisfies the condition
\begin{equation*}
\partial _{i}\Omega -\left( w_{i}+\zeta _{i}\right) \partial _{\nu }\Omega =0
\end{equation*}%
for any additional deformation functions $\zeta _{i}(\theta ,\nu )$ and the
coefficients
\begin{equation}
g_{1}=1,g_{2}=g_{2[0]}q^{[v]}\left( \theta \right) ,\ h_{3,4}=\eta
_{3,4}(\theta ,\nu )h_{3,4[0]},w_{i}(\theta ,\nu ),n_{i}(\theta ,\nu )
\label{data5}
\end{equation}%
satisfy the equations (\ref{ricci1a})--(\ref{ricci4a}). The procedure of
constructing such exact solutions is very similar to the considered in
subsection 4.1 for black ellipsoids. For anholonomic ellipsoidal shapes
(they are characterized by nontrivial anholonomy coefficients (\ref{anh})
and respectively induced noncommutative symmetries) we have to put as
''boundary'' condition in integrals of type (\ref{auxf4}) just to have $%
n_{1}=0,\ n_{2}=n_{2[0]}(\theta )=2R_{0}\cos \theta $ from data (\ref{dataes}%
) in the limit when dependence on ''anisotropic'' variable $\nu $ vanishes.
The functions $w_{i}(\theta ,\nu )$ and $n_{i}(\theta ,\nu )$ must be
subjected to additional constraints if we wont to construct ellipsoidal
shape configurations with zero anholonomically induced torsion (\ref{torsion}%
) and N--connection curvature, $\Omega _{jk}^{a}=\delta _{k}N_{j}^{a}-\delta
_{j}N_{k}^{a}=0.$

b) The simplest way to construct an ellipsoidal shape configuration induced
by anisotropic cosmological constant is to find data (\ref{data5}) solving
the equations (\ref{eq17}) following the procedure defined in subsection
4.2. \ We note that we can solve the equation (\ref{eqaux1}) for $%
g_{2}=g_{2}\left( \theta \right) =\sin ^{2}\theta $ with $q^{[v]}\left(
\theta \right) =1$ if $\lambda _{\lbrack h]0}=1/2,$ see solution (\ref{aux2p}%
) with $\xi \rightarrow \theta .$ For simplicity, we can consider that $%
\lambda _{\lbrack v]}=0.$ Such type configurations contain, in general,
anholonomically induced torsion.

We conclude, that by using the anholonomic frame method we can generate
ellipsoidal shapes (in general, with nontrivial pollarized cosmological
constants and induced torsions). Such solutions are similar to corresponding
rotation configurations in general relativity with rigidly rotating
perfect-fluid sources. The rough analogy consists in the fact that by
certain frame constraints induced by off--diagonal metric terms we can model
gravitational--matter like metrics. In previous section we proved the
stability of black ellipsoids for small excentricities. Similar
investigations for ellipsoidal shapes is a task for future (because the
shapes could be with arbitrary excetricity). In Ref. \cite{es}, there were
discussed points of matchings of locally rotationally symmetric spacetimes
to Taub--NUT metrics. We emphasize that this topic was also specifically
elaborated by using anholonomic frame transforms in Refs. \cite{vspd1}.

\subsection{Generalization of Canfora--Schmidt solutions}

In general, the solutions generated by anholonomic transforms cannot be
reduced to a diagonal transform only by coordinate transforms (this is
stated in our previous works \cite%
{v,v1,vsbd,vspd1,vncs,vth,velp1,velp2,vncfg,vels,vncs}, see also Refs. \cite%
{vd} for modelling Finsler like geometries in (pseudo) Riemannian
spacetimes). \ We descuss here how 4D off--diagonal ansatz (\ref{ansatzc4})
generalize the solutions obtained in Ref. \cite{cans} by a corresponding
parametrization of coordinates as $x^{1}=x,x^{2}=t,y^{3}=\nu =y$ and $%
y^{4}=p.$ If we consider for (\ref{ansatzc4}) (equivalently, for (\ref%
{dmetric4}) ) the non-trivial data%
\begin{eqnarray}
g_{1} &=&g_{1[0]}=1,\ g_{2}=g_{2[0]}(x^{1})=-B\left( x\right) P(x)^{2}-C(x),
\notag \\
h_{3} &=&h_{3[0]}(x^{1})=A\left( x\right) >0,\ h_{4}=h_{4[0]}(x^{1})=B\left(
x\right) ,  \notag \\
w_{i} &=&0,n_{1}=0,n_{2}=n_{2[0]}(x^{1})=P(x)/B\left( x\right)  \label{data6}
\end{eqnarray}%
we obtain just the ansatz (12) from Ref. \cite{cans} (in this subsection we
use a different label for coordinates) which, for instance, for $%
B+C=2,B-C=\ln |x|,P=-1/(B-C)$ with $e^{-1}<\sqrt{|x|}<e$ for a constant $e,$
defines an exact 4D solution of the Einstein equation (see metric (27) from %
\cite{cans}). By introducing 'polarization' functions $\eta _{k}=\eta
_{k}(x^{i})$ [when $i,k,...=1,2]$ and $\eta _{a}=\eta _{a}(x^{i},\nu )$
[when $a,b,...=3,4$] we can generalize the data (\ref{data6}) as to have
\begin{equation*}
g_{k}(x^{i})=\eta _{k}(x^{i})g_{k[0]},\ h_{a}(x^{i},\nu )=\eta
_{a}(x^{i},\nu )h_{a[0]}
\end{equation*}%
and certain nontrivial values $w_{i}=w_{i}(x^{i},\nu )$ and $%
n_{i}=n_{i}(x^{i},\nu )$ solving the Einstein equations with anholonomic
variables (\ref{ricci1a})--(\ref{ricci4a}). We can easy find new classes of
exact solutions, for instance, for $\eta _{1}=1$ and $\eta _{2}=\eta
_{2}(x^{1}).$ In this case $g_{1}=1$ and the function $g_{2}(x^{1}) $ is any
solution of the equation
\begin{equation}
g_{2}^{\bullet \bullet }-\frac{(g_{2}^{\bullet })^{2}}{2g_{2}}=0
\label{partic1}
\end{equation}%
(see equation (\ref{eqaux1}) for $\lambda _{\lbrack v]}=0),$ $g_{2}^{\bullet
}=\partial g_{2}/\partial x^{1}$ which is solved as a particular case if $%
g_{2}=(x^{1})^{2}.$ This impose certain conditions on $\eta _{2}(x^{1})$ if
we wont to take $g_{2[0]}(x^{1})$ just as in (\ref{data6}). For more general
solutions with arbitrary $\eta _{k}(x^{i}),$ we have to take solutions of
equation (\ref{ricci1a}) and not of a particular case like (\ref{partic1}).

We can generate solutions of (\ref{ricci2a}) for any $\eta _{a}(x^{i},\nu )$
satisfying the condition (\ref{conda}), $\sqrt{|\eta _{3}|}=\eta _{0}\left(
\sqrt{|\eta _{4}|}\right) ^{\ast },\eta _{0}=const.$ For instance, we can
take arbitrary $\eta _{4}$ and using elementary derivations with $\eta
_{4}^{\ast }=\partial \eta _{4}/\partial \nu $ and a nonzero constant $\eta
_{0},$ to define $\sqrt{|\eta _{3}|}.$ For the vacuum solutions, we can put $%
w_{i}=0$ because $\beta =\alpha _{i}=0$ (see \ formulas (\ref{ricci2a}) and (%
\ref{abc})). In this case the solutions of (\ref{ricci3a}) are trivial.
Having defined $\eta _{a}(x^{i},\nu )$ we can integrate directly the
equation (\ref{ricci4a}) and find $n_{i}(x^{i},\nu )$ like in formula (\ref%
{auxf4}) with fixed value $\varepsilon =1$ and considering dependence on all
holonomic variables,
\begin{eqnarray}
n_{i}\left( x^{k},\nu \right) &=&n_{i[1]}\left( x^{k}\right) +n_{i[2]}\left(
x^{k}\right) \int d\nu \ ~\eta _{3}\left( x^{k},\nu \right) /\left( \sqrt{%
|\eta _{4}\left( x^{k},\nu \right) |}\right) ^{3},\eta _{4}^{\ast }\neq 0;
\notag \\
&=&n_{i[1]}\left( x^{k}\right) +n_{i[2]}\left( x^{k}\right) \int d\nu \ \eta
_{3}\left( x^{k},\nu \right) ,\eta _{4}^{\ast }=0;  \label{formau} \\
&=&n_{i[1]}\left( x^{k}\right) +n_{i[2]}\left( x^{k}\right) \int d\nu
/\left( \sqrt{|\eta _{4}\left( x^{k},\nu \right) |}\right) ^{3},\eta
_{3}^{\ast }=0.  \notag
\end{eqnarray}%
These values will generalize the data (\ref{data6}) if we identify $%
n_{1[1]}\left( x^{k}\right) =0$ and $n_{1[1]}\left( x^{k}\right)
=n_{2[0]}(x^{1})=P(x)/B\left( x\right) .$ The solutions with vanishing
induced torsions and zero nonlinear connection curvatures are to be selected
by choosing $n_{i}\left( x^{k},\nu \right) $ and $\eta _{3}\left( x^{k},\nu
\right) $ (or $\eta _{4}\left( x^{k},\nu \right) )$ as to reduce the
canonical connection (\ref{dcon}) to the Levi--Civita connection (as we
discussed in the end of Section 2).

The solution defined by the data (\ref{data6}) is compared in Ref. \cite%
{cans} with the Kasner diagonal solution which define the simplest models of
anisotropic cosmology. The metrics obtained by F. Canfora and H.-J. Schmidt
(CS) is generic off--diagonal and can not written in diagonal form by
coordinate transforms. We illustrated that the CS metrics can be effectively
diagonalized with respect to N--adapted anholonomic frames (like a more
general ansatz (\ref{ansatzc4}) can be reduced to (\ref{dmetric4})) and that
by anholonomic frame transforms of the CS metric we can generate new classes
of generic off--diagonal  solutions. Such spacetimes may describe certain
models of anisotropic and/or inhomogeneous cosmologies (see, for instance,
Refs. \cite{vd} were we considered a model of Friedman--Robertson--Walker
metric with ellipsoidal symmetry). The anholonomic generalizations of CS\
metrics are with nontrivial noncommutative symmetry because the anholonomy
coefficients (\ref{anh}) (see also (\ref{anhb})) are not zero being defined
by nontrivial values (\ref{formau}).

\section{Outlook and Conclusions}

The work is devoted to investigation of a new class of exact solutions in
metric--affine and string gravity describing static back rotoid (ellipsoid)
\ and shape configurations possessing hidden noncommutative symmetries.
There are generated also certain generic off--diagonal cosmological metrics.

We consider small, with nonlinear gravitational polarization, static
deformations of the Schwarschild black hole solution (in particular cases,
to some resolution ellipsoid like configurations) preserving the horizon and
geodesic behaviour but slightly deforming the spherical constructions. It
was proved that there are such parameters of the exact solutions of the
Einstein equations defined by off--diagonal metrics with ellipsoid symmetry
constructed in Refs. \cite{v,v1,vth,velp1,velp2,vels} as the vacuum
solutions positively define static ellipsoid black hole configurations.

We illustrate that the new class of static ellipsoidal black hole solutions
posses some similarities with the Reissner--Nordstrom metric if the metric's
coefficients are defined with respect to correspondingly adapted anholonomic
frames. \ The parameter of ellipsoidal deformation results in an effective
electromagnetic charge induced by off--diagonal vacuum gravitational
interactions. We note that effective electromagnetic charges and
Reissner--Nordstrom metrics induced by interactions in the bulk of extra
dimension gravity were considered in brane gravity \cite{maartens}. In our
works we proved that such Reissner--Nordstrom like ellipsoid black hole
configurations may be constructed even in the framework of vacuum Einstein
gravity. It should be emphasized that the static ellipsoid black holes
posses spherical topology and satisfy the principle of topological
censorship \cite{haw1}. Such solutions are also compatible with the black
hole uniqueness theorems \cite{ut}. In the asymptotical limits at least for
a very small eccentricity such black ellipsoid metrics transform into the
usual Schwarzschild one. We have proved that the stability of static
ellipsoid black holes can be proved similarly by considering small
perturbations of the spherical black holes \cite{velp1,velp2} even the
solutions are extended to certain classes of spacetimes with anisotropically
polarized cosmological constants. (On the stability of the Schwarzschild
solution see details in Ref. \cite{chan}.)

The off--diagonal metric coefficients induce a specific spacetime distorsion
comparing to the solutions with metrics diagonalizable by coordinate
transforms. So, it is necessary to compare the off--diagonal ellipsoidal
metrics with those describing the distorted diagonal black hole solutions
(see the vacuum case in Refs. \cite{ms} and an extension to the case of
non--vanishing electric fields \cite{fk}). For the ellipsoidal cases, the
distorsion of spacetime can be of vacuum origin caused by some anisotropies
(anholonomic constraints) related to off--diagonal terms. In the case of
''pure diagonal'' distorsions such effects follow from the fact that the
vacuum Einstein equations are not satisfied in some regions because of
presence of matter.

The off--diagonal gravity may model some gravity--matter like interactions
like in Kaluza--Klein theory (for some particular configurations and
topological compactifications) but, in general, the off--diagonal vacuum
gravitational dynamics can not be associated to any effective matter
dynamics. So, we may consider that the anholonomic ellipsoidal deformations
of the Schwarzschild metric are some kind of anisotropic off--diagonal
distorsions modeled by certain vacuum gravitational fields with the
distorsion parameteres (equivalently, vacuum gravitational polarizations)
depending both on radial and angular coordinates.

There is a common property that, in general, both classes of off--diagonal
anisotropic and ''pure'' diagonal distorsions (like in Refs. \cite{ms})
result in solutions which are not asymptotically flat. However, it is
possible to find asymptotically flat extensions even for ellipsoidal
configurations by introducing the corresponding off--diagonal terms (the
asymptotic conditions for the diagonal distorsions are discussed in Ref. %
\cite{fk}; to satisfy such conditions one has to include some additional
matter fields in the exterior portion of spacetime).

We analyzed the conditions when the anholonomic frame method can model
ellipsoid shape configurations. It was demonstrated that the off--diagonal
metric terms and respectively associated nonlinear connection coefficients
may model ellipsoidal shapes being similar to those derived from solutions
with rotating perfect fluids (roughly speaking, a corresponding frame
anholnomy/ anisotropy may result in modeling of specific matter interactions
but with polarizations of constants, metric coefficients and related frames).

In order to point to some possible observable effects, we note that for the
ellipsoidal metrics with the Schwarzschild asymptotics, the ellipsoidal
character could result in some observational effects in the vicinity of the
horizon (for instance, scattering of particles on a static ellipsoid; we can
compute anisotropic matter accretion effects on an ellipsoidal black hole
put in the center of a galactic being of ellipsoidal or another
configuration). A point of further investigations could be the anisotropic
ellipsoidal collapse when both the matter and spacetime are of ellipsoidal
generic off--diagonal symmetry and/or shape configurations (former
theoretical and computational investigations were performed only for rotoids
with anisotropic matter and particular classes of perturbations of the
Schwarzshild solutions \cite{st}). For very small eccentricities, we may not
have any observable effects like perihelion shift or light bending if we
restrict our investigations only to the Schwarzshild--Newton asymptotics.

We present some discussion on mechanics and thermodynamics of ellipsoidal
black holes. For the static black ellipsoids with flat asymptotics, we can
compute the area of the ellipsoidal horizon, associate an entropy and
develop a corresponding black ellipsoid thermodynamics. This can be done
even for stable black torus configurations. But this is a very rough
approximation because, in general, we are dealing with off--diagonal metrics
depending anisotropically on two/three coordinates. Such solutions are with
anholonomically deformed Killing horizons and should be described by a
thermodynamics (in general, both non-equilibrium and irreversible) of black
ellipsoids self--consistently embedded into an off--diagonal anisotropic
gravitational vacuum. This is a ground for numerous new conceptual issues to
be developed and related to anisotropic black holes and the anisotropic
kinetics and thermodynamics \cite{v1} as well to a framework of isolated
anisotropic horizons \cite{asht} which is a matter of our further
investigations. As an example of a such new concept, we point to a
noncommutative dynamics which can be associated to black ellipsoids.

We emphasize that it is a remarkable fact that, in spite of appearance
complexity, the perturbations of static off--diagonal vacuum gravitational
configurations are governed by similar types of equations as for diagonal
holonomic solutions. Perhaps in a similar manner (as a future development of
this work) by using locally adapted ''N--elongated'' partial derivatives we
can prove stability of very different classes of exact solutions with
ellipsoid, toroidal, dilaton and spinor--soliton symmetries constructed in
Refs. \cite{v,v1,vth,velp1,velp2,vels}. The origin of this mystery is
located in the fact that by anholnomic transforms we effectively
diagonalized the off--diagonal metrics by ''elongating'' some partial
derivatives. This way the type of equations governing the perturbations is
preserved but, for small deformations, the systems of linear equations for
fluctuations became ''slightly'' nondiagonal and with certain tetradic
modifications of partial derivatives and differentials.

It is known that in details the question of relating the particular
integrals of such systems associated to systems of linear differential
equations is investigated in Ref. \cite{chan}. For anholonomic
configurations, one holds the same relations between the potentials $%
\widetilde{V}^{(\eta )}$ and $V^{(-)}$ and wave functions $Z^{(\eta )}$ and $%
Z_{(A)}^{(+)}$ with that difference that the physical values and formulas
where polarized by some anisotropy functions $\eta _{3}(r,\theta ,\varphi
),\Omega (r,\varphi ),q(r),\eta (r,\varphi ),$ $w_{1}(r,\varphi )$ and $%
n_{1}(r,\varphi )$ and deformed on a small parameter $\varepsilon .$ It is
not clear that a similar procedure could be applied in general for proofs of
stability of ellipsoidal shapes but it would be true for small deformations
from a supposed to be stable primordial configuration.

We conclude that there are static black ellipsoid vacuum configurations as
well induced by nontrivially polarized cosmological constants which are
stable with respect to one dimensional perturbations, axial and/or polar
ones, governed by solutions of the corresponding one--dimensional
Schrodinger equations. The problem of stability of such objects with respect
to two, or three, dimensional perturbations, and the possibility of modeling
such perturbations in the framework of a two--, or three--, dimensional
inverse scattering problem is a topic of our further investigations. The
most important problem to be solved is to find a geometrical interpretation
for the anholonomic Schrodinger mechanics of stability to the anholonomic
frame method and to see if we can extend the approach at least to the two
dimensional scatering equations.


\subsection*{Acknowledgements}

~~ The work is partially supported by a NATO/Portugal fellowship at CENTRA,
Instituto Superior Tecnico, Lisbon. The author is grateful for support to
the organizes of the International Congress of Mathematical Physics, ICMP
2003, and of the Satelite\ Meeting OPORTO 2003. He would like to thank J.
Zsigrai for valuable discussions.



\end{document}